\documentclass[letterpaper]{article}

\usepackage[T1]{fontenc}
\usepackage{amssymb}
\usepackage{braket}
\usepackage{algorithm}
\usepackage{algpseudocode}
\usepackage{placeins}

\usepackage{geometry}
\usepackage{setspace}
\usepackage{booktabs}

\usepackage[
    colorlinks=true,
    citecolor=red,
    linkcolor=blue,
    urlcolor=blue
]{hyperref}

\usepackage[
    backend=bibtex,
    style=numeric-comp,
    sorting=none
]{biblatex}

\usepackage{graphicx}
\usepackage{float}
\newfloat{scheme}{htbp}{los}
\floatname{scheme}{Scheme}
\floatname{chart}{Chart}
\newfloat{graph}{htbp}{loh}

\usepackage{chemformula} 
\usepackage[version = 4]{mhchem} 
\DeclareMathOperator*{\argmin}{arg\,min}

\usepackage{authblk}
\author[1]{Jonas Beck}
\author[1]{Michele Nottoli}
\author[2]{Tommaso Nottoli}\author[2]{Luca Melega}
\affil[1]{Institute of Applied Analysis and Numerical Simulation, University of Stuttgart, Pfaffenwaldring 57, 70569 Stuttgart, Germany}
\author[2]{Filippo Lipparini}
\affil[2]{Dipartimento di Chimica e Chimica Industriale
Via Moruzzi 13, 56124 Pisa, Italy}
\author[1]{Benjamin Stamm}

\title{Coupled Cluster Amplitude Extrapolation in Geometry Optimization}
\date{*Email: jonas.beck@ians.uni-stuttgart.de}

\begin{document}

\maketitle

\begin{abstract}
  In this work, we present an amplitude extrapolation ansatz for geometry optimization that uses calculations from previous steps to construct an initial guess for the quasi-Newton solver at each new geometry, significantly accelerating convergence of the coupled-cluster and Lambda solver alike. The ansatz incorporates a necessary amplitude transformation to mitigate discontinuities associated with orbital energy crossings at the Hartree–Fock level of theory, while remaining computational and memory efficient.
  The method is tested
  for a diverse set of medium-sized molecules to demonstrate a systematic and significant reduction in the number of coupled-cluster iterations compared to the conventional method.
\end{abstract}





\section{Introduction}
The calculation of molecular equilibrium structures is one of the most fundamental tasks in computational quantum chemistry. For small- to medium-sized molecular systems, that are typically represented by one or few conformers, determining their equilibrium geometry is not only relevant \emph{per se}, but also the common starting point for subsequent computational investigations, including the calculation of spectroscopic and molecular properties.
Within the Born-Oppenheimer approximation, the equilibrium structures can be seen as minimum-energy points on the molecular potential energy surface (PES), and therefore can be found by solving a numerical optimization problem, commonly known as geometry optimization. The most popular techniques in use are based on a quasi-Newton strategy~\cite{Fletcher1999,schlegel2011geometry}, an iterative procedure that requires, at each geometry, the calculation of the molecular energy and its analytical gradient~\cite{helgaker_properties,Pulay1977grad,schlegel2011geometry}.

Geometry optimizations, therefore, require repeated energy and gradient evaluations, which, in turn, require the solution to the equations appropriate for the level of theory set for the problem. For non-variational methods, such as Coupled Cluster (CC) methods~\cite{shavitt2009many}, the calculation of the gradients requires the solution to an additional set of equations, known as $\Lambda$ equations~\cite{adamowicz1984lambda,scheiner1987analytic,salter1989analytic,gauss1991coupled,GAUSS1991207rohf,gauss1991qrhf,scuseria1991analytic,lee1991analytic,WATTS19921open,watts93,Scuseria88, Gauss00,gauss2002ccsdt,hald2003lagrangian,kallay2003analytic,harding2008parallel}, required to set up an appropriate Lagrangian~\cite{helgaker1982simple,koch1990lagrangian}, that can then be differentiated efficiently. This makes geometry optimization a computationally intensive task. Given that the quantum mechanical equations that define the various levels of theory are typically solved iteratively, using information from previous geometry optimization steps to provide an accurate guess to such an iterative procedure is an attractive way to reduce the overall computational cost of the optimization. 

Recently, Polack et al. proposed an extrapolation method aiming to get sophisticated initial guess strategies for Hartree Fock (HF) and Density Functional Theory (DFT) in the context of Born–Oppenheimer molecular dynamics (BOMD)~\cite{Polack17102020, 10.1021/acs.jctc.1c00751, 10.1021/acs.jpclett.3c02098, 10.1063/5.0323068}. These methods were subsequently extended to the context of geometry optimization~\cite{10.1021/acs.jctc.4c01417}. In a Taylor-based approach, the calculated quantities at the previous geometries are used to build an ansatz that serves as an initial guess, incorporating information about the new geometry via a so-called molecular descriptor.

In this contribution, we propose a conceptually similar strategy that can be applied to Coupled Cluster optimizations. CC methods have proven highly reliable for the accurate computation of molecular properties and structures, as they are systematically improvable and size extensive. For most applications, CC with singles and doubles (CCSD) provides a good trade-off between cost and accuracy~\cite{10.1063/1.443164}. It involves determining the CC amplitudes by solving a large system of nonlinear equations using an iterative quasi-Newton method~\cite{1212166}. The main computational bottleneck lies in the evaluation of the residual at each Newton step, with a formal $\mathcal{O}(N^6)$ computational scaling. This high computational cost restricts the application of CCSD to medium-sized systems. The calculation of CC analytical gradients is about twice as expensive as the energy calculation, since the $\Lambda$ equations have the same scaling, and overall similar computational cost, as the amplitude equations.
For this reason, various strategies were proposed in the literature to reduce the computational cost and memory requirements while preserving accuracy, for instance exploiting the local nature of dynamic electron correlation~\cite{hampel1996local,schuetz2001,rauhut2001analytical,werner2011efficient,wei2009local,neese2009efficient,neese2009efficient2,datta2016analytic,kjaergaard2017divide,rolik2011general,riplinger2013dlpno,guo2018dlpno,ma2018explicitly} or using rank-reducing techniques~\cite{epifanovsky2013general,peng2017low,Nottoli2023,deprince2013accuracy,Pedersen2024cd,parrish2012hyper,parrish2019rank,hohenstein2022rank,bozkaya2016analytic,bozkaya2017triples,feng2019implementation,schnack2022efficient,melega2026novel}. 

From a different, yet complementary, perspective, fewer attempts have been made to provide a better guess for solving the CC equations, thus reducing the overall number of iterations during a geometry optimization procedure. Therefore, generating a high-quality initial guess remains a primary objective, where a standard comparison point is represented by the guess amplitudes being provided by second-order Møller–Plesset Perturbation Theory (MP2)~\cite{mller_plesset_1934, helgakar}.

It is important here to notice that the extrapolation methods developed for HF and DFT cannot be directly extended to CC. In HF and DFT, the natural object to extrapolate is the density matrix, which is invariant for changes in the orbitals that only mix occupied with occupied ones. 
For Coupled Cluster theory, the quantities that need to be extrapolated are the Coupled Cluster amplitudes and Lagrange multipliers, that in turn, explicitly depend on both occupied and virtual orbitals and, even worse, on their order. This is a major complication, because the extrapolation method developed requires the extrapolated quantities to be smooth: given that at different geometries the occupied and virtual orbitals can undergo energy crossings, the CC amplitudes and Lagrange multipliers become discontinuous. 
An approach to circumvent these problems was proposed by using a consistently smooth set of molecular orbitals \cite{10.1063/5.0141145}.
This strategy however comes with a drawback,
using an alternative set of orbitals directly is generally not advisable because canonical orbitals ensure that the approximate Jacobian, used as a preconditioner in the coupled cluster quasi-Newton method, remains strictly diagonal. An alternative approach, which was proposed by some of us \cite{beck2026regularityinterpolationcoupledcluster}, addresses the non-smoothness of the amplitudes by using an orbital transformation which is computationally and memory efficient. The transformation was then used to perform the interpolation of the amplitudes for a toy model inspired by BOMD. This transformation is leveraged in this work to apply the extrapolation schemes proposed in \cite{10.1021/acs.jctc.4c01417} to both the CC and Lambda amplitudes. The transformation effectively moves the extrapolation process to the atomic orbital basis, where no issues due to orbital crossings are present, without ever exceeding the memory requirements of $\mathcal{O}(V^2O^2)$, where $V$ denotes the number of virtual orbitals and $O$ the number of occupied orbitals.

The article is structured as follows. In the Theory Section, we outline the necessary theoretical background, starting from a brief review of the coupled cluster method and the Lagrange formulation.
This is followed by an introduction to the extrapolation approach, which is used to obtain good initial guesses for the iterative coupled-cluster solver and for the iterative scheme to solve for the $\Lambda$-amplitudes. The Theory Section concludes with a discussion of the amplitude transformation technique employed to quasi-smoothen the amplitudes and a demonstration that this transformation ansatz is essential for guaranteeing reliable and accurate results. In the Results Section, we first perform a parameter calibration on a set of medium-sized molecules to identify the extrapolation parameters that yield the best performance for the majority of the tested systems. Subsequently, we provide extensive benchmark data off a larger set of molecules to demonstrate that using the extrapolants as initial guesses systematically reduces the required number of Newton steps for both the coupled-cluster and the Lambda solver.

\section{Theory}
\label{sec:theory}

\subsection{Coupled cluster theory}
The CC wave function ansatz is written as
\begin{align}\label{exp_param}
    \ket{\Psi}= \mathrm{e}^{T}\ket{0},
\end{align}
where $\ket{0}$ denotes the Hartree-Fock reference determinant, namely the mean-field solution, and $T$ a suitable cluster operator. Adopting the usual index convention that $i,j,k\ldots$ refer to occupied orbitals, $a,b,c,\ldots $ to virtual orbitals, $p,q,r,\ldots $ to generic orbitals and $\mu, \nu,\sigma,\ldots $ to atomic orbitals the manifold of admissible cluster operators can be written as
\begin{align}\label{cluster_op}
    T = T_1 + T_2 + T_3 \cdots
\end{align}
with
\begin{equation}\label{eq:Tn_operator}
    T_n = \frac{1}{(n!)^2}\sum_{\substack{ab\cdots \\ ij\cdots}}t_{ij\cdots}^{ab\cdots}\{\hat{a}^\dagger \hat{i}\hat{b}^\dagger \hat{j}\cdots\}
\end{equation}
as a generic cluster operator. In Eq.\eqref{eq:Tn_operator}, $t_{ij\cdots}^{ab\cdots}$ are the CC T-amplitudes, and $\{\hat{a}^\dagger \hat{i}\hat{b}^\dagger \hat{j}\cdots\}$ the normal-ordered string of excitation operators. It is well-known that the exponential parametrization \eqref{exp_param} corresponds to the exact FCI solution if all cluster operators are considered~\cite{rohwedder1, Schneider2009, COESTER1960477}. In practice one only considers cluster operators up to excitation level $n$ by restricting the admissible operators $T$ in \eqref{cluster_op}. In this contribution, we focus on the CCSD method, even if the results and methods discussed later are valid for all coupled cluster truncation levels.

In this case, the cluster operator reads
\begin{equation}
    T = \sum_{ai}t_i^a\{\hat{a}^\dagger \hat{i}\} + \frac{1}{4}\sum_{abij}t_{ij}^{ab}\{\hat{a}^\dagger \hat{i}\hat{b}^\dagger \hat{j}\}
\end{equation}
and the optimal T-amplitudes are obtained by solving a set of projected nonlinear equations
\begin{align}\label{CCSD_equations}
     \langle \Phi^{a}_{i}| \bar{H} |0 \rangle =0,\quad \langle \Phi^{ab}_{ij}| \bar{H}| 0 \rangle =0,
\end{align}
where $\bar{H}:=\mathrm{e}^{-T}H\mathrm{e}^T$ is the similarity transformed electronic Hamiltonian, and $\ket{\Phi_i^a}$ and $\ket{\Phi_{ij}^{ab}}$ are singly and doubly excited Slater determinants, respectively. After solving the system of nonlinear equations \eqref{CCSD_equations}, the corresponding energy is obtained via 
\begin{align}\label{CCSD_energy}
    \mathcal{E}_{CC}=\langle 0|\bar{H}|0\rangle \text{.}
\end{align}

\subsection{Nuclear gradients and the Lagrange equations}
Since the CC method is non variational, the Lagrangian formalism is commonly employed to enable the efficient evaluation of energy gradients \cite{helgaker1988analytical,koch1990lagrangian}. By defining the de-excitation cluster operator
\begin{equation}
    \Lambda = \sum_{ai}\lambda_i^a\{\hat{i}^\dagger \hat{a}\}+\frac{1}{4}\sum_{abij}\lambda_{ab}^{ij}\{\hat{i}^\dagger\hat{a}\hat{j}^\dagger\hat{b}\},
\end{equation}
the CCSD Lagrangian can be written as
\begin{equation}\label{Lagrangian}
    \mathcal{L}(t,\lambda, Z, I):= \langle 0| (1+\Lambda)\bar{H}|0\rangle + 2\sum_{ai}Z_{ai}f_{ai} + \sum_{pq}I_{pq}\left(\sum_{\mu\nu}C_{\mu p}S_{\mu\nu}C_{\nu q} -\delta_{pq}\right).
\end{equation}

with $t$ and $\lambda$ being the vectors containing the T and $\Lambda$ amplitudes, respectively. $Z_{ai}$ and $I_{pq}$ are Lagrange multipliers used to enforce the Brillouin condition and the orthonormality of the molecular orbitals ($C_{\mu p}$), respectively. Stationarity of the Lagrangian with respect to the $\Lambda$-amplitudes gives exactly the system of equations required to get the T-amplitudes (cf. Eqs.~\eqref{CCSD_equations}), while stationarity with respect to T provides the expressions for the $\Lambda$ equations
\begin{align}\label{Lambda_eq}
    \frac{\partial \mathcal{L}}{\partial t^a_i}=\langle \Phi^{a}_{i}|(1+\Lambda) [\bar{H},\{\hat{a}^\dagger\hat{i}\}] |0 \rangle =0,\, \frac{\partial \mathcal{L}}{\partial t^{ab}_{ij}}=\langle \Phi^{ab}_{ij}|(1+\Lambda) [\bar{H},\{\hat{a}^\dagger\hat{i}\hat{b}^\dagger\hat{j}\}] |0 \rangle =0.
\end{align}
Enforcing stationarity with respect to orbital rotations then yields expressions for the $I_{pq}$ and $Z_{ai}$ multipliers, with the latter giving rise to the so-called Z-vector equations.

With all parameters determined, the expression for the energy is written in the usual form
\begin{equation}
    \frac{d\mathcal{E}_{CC}}{dx} = \sum_{\mu \nu} D_{\mu \nu} h_{\mu \nu}^x + \frac{1}{2}\sum_{\mu \nu \rho \sigma} \Gamma_{\mu \rho \nu \sigma} (\mu \nu \vert \rho \sigma)^x + \sum_{\mu \nu} I_{\mu \nu} S_{\mu \nu}^x,
\end{equation}
where $D$ and $\Gamma$ are the relaxed one- and two-body density matrices written in the AO basis.

\subsection{Extrapolation ansatz} The coupled cluster equations~\eqref{CCSD_equations} and the Lambda equations~\eqref{Lambda_eq} are both solved via a quasi-Newton method. Each iteration requires the evaluation of a residual, which has the steep scaling of $\mathcal{O}(V^4O^2)$, which corresponds to a large computational cost for medium sized molecules.
This large cost can be reduced by reducing the number of necessary quasi-Newton iterations with a good initial guess. For that, we adapt the extrapolation approach used in previous works on DFT \cite{Polack17102020, 10.1021/acs.jctc.1c00751, 10.1021/acs.jctc.4c01417}.

Given a set of previous geometries $\lbrace r_1,\ldots,r_q \rbrace$ and a new target geometry $r^*$, the extrapolation scheme tries to find $\alpha_1,\ldots,\alpha_q$ such that the ansatz
\begin{align}\label{extrap_approx}
    \tilde{t}= \sum_{j=1}^q\alpha_j t(r_j),
\end{align}
is a good approximation to the true amplitudes $t(r^*)$. As the function that maps the molecular geometry to the amplitudes is highly non-linear and not suited for the extrapolation, we use a suitable molecular descriptor $d$, which is assumed to allow for the decomposition $t(r)=g(d(r))$ for some unknown function $g$, which is however easier to approximate.

Assuming the existence of a smooth function $g$ with respect to $t$ and $d$, and by using the new geometry~$r^*$ as a reference, one can for any $j\le q$ Taylor expand
\begin{align}\label{Taylor_t}
    t(r_j)= t(r^*)+Dg(d^*)\Delta d_j+ \mathcal{O}(\|\Delta d_j\|^2),
\end{align}
where $Dg(d^*)$ is the Jacobian of $g$ with respect to $d$ at point $d^*:=d(r^*)$, and $\Delta d_j= d(x_j)-d^*$. Using the function evaluations at the previous geometries $\lbrace t(r_1),\ldots,t(r_q) \rbrace$ and given a set of real coefficients $\alpha_1,\ldots,\alpha_q$ with $\sum_j \alpha_j =1$ the difference of \eqref{extrap_approx} to the true solution can be written as 
\begin{align*}
    t(r^*)-\tilde{t}&=t(r^*)-\sum_{j=1}^q \alpha_j \big(t(r^*)+Dg(d^*)\Delta d_j+ \mathcal{O}(\|\Delta d_j\|^2)\big)\\
    &= -Dg(d^*)\sum^q_{j=1}\alpha_j \Delta d_j + \underbrace{\sum^{q}_{j=1} \alpha_j \mathcal{O}( \|\Delta d_j\|^2)}_{=:\,\Delta D}.
\end{align*}
Which yields, due to $\sum_j\alpha_j=1$, 
\begin{align}\label{error_estimate}
    \|t(r^*)-\tilde{t}\| \le \|Dg(d^*)\| \| d(r^*)-\sum^q_{j=1}\alpha_j d(r_j)\|+ \Delta D.
\end{align}
Hence, if the descriptor values are close, i.e. $\Delta D \approx 0$ 
for the new geometry $r^*$, the error reduces to picking the optimal choice of coefficients, which is obtained by solving the constrained minimization problem
\begin{align}\label{least_sq_desc_fit}
\alpha^*:=\argmin_{\alpha_1,\ldots,\alpha_q}\big\|d(r^*)-\sum^{q}_{j=1}\alpha_j d(r_j)\big\|^2,\quad\sum^q_{j=1}\alpha_j=1.
\end{align}
The error estimate \eqref{error_estimate} splits in two fundamentally different error contributions. The first term corresponds to the \textit{subspace error} in the descriptor space, with a conditioning factor depending on the derivative of the unknown function $g$, and the second term $\Delta D$ corresponds to the sum of the \textit{Taylor errors}. Although the subspace error generally decreases when considering as many previous descriptor values as possible, there are two problems with this strategy. First, the Taylor error, as a sum, increases if more descriptor evaluations are included. Secondly, the minimization problem \eqref{least_sq_desc_fit} can become poorly conditioned if linear dependencies arise. 

The first problem worsens when considering descriptor evaluations on geometries that are far from the new geometry, which usually correspond to the oldest steps in a geometry optimization. Conversely, the problem of linear dependencies typically arises when descriptor values are too close to one another. To address the latter issue, the least squares minimization \eqref{least_sq_desc_fit} is replaced by a regularized minimization problem \cite{doi:10.1137/1021044, 481c956f-fc76-3a47-912c-132550d4ccb6}. To mitigate the Taylor error, a moving window approach can be employed that considers only the most recent $q$ steps.

Therefore, 
the extrapolation method allows for different choices of extrapolation parameters
\begin{align}\label{extrap_parameter}
    \theta=(d,q, \varepsilon),
\end{align}
where $d$ is the descriptor choice, $q$ the number of previous snapshots to consider, i.e. the size of the moving window, and $\varepsilon$ the choice of the Tikhonov constant for regularization.
\subsection{Amplitude transformation} The extrapolation method, discussed in the previous section, heavily relies on the regularity of the amplitudes with respect to the nuclear coordinates. It is well known that on the Hartree-Fock level of theory orbital energy crossings may occur, which result in reorderings of the coefficient matrix in the canonical orbital basis. This leads to discontinuities that are ultimately carried over to the coupled cluster amplitudes, making them not directly suitable for any extrapolation strategy.

Recently, some of the authors introduced an amplitude transformation to efficiently address these issues~\cite{beck2026regularityinterpolationcoupledcluster}. A similar transformational ansatz was used in \cite{https://doi.org/10.1002/jcc.70473}. The amplitude tensor transformation with respect to geometry $r'$ and $r$ has the form of 
\begin{align}\label{Transformation}
    \mathbf{L}(r',r)t(r):=\left[\mathbf{A}^{-1}(r')\mathbf{A}(r)\right]t(r).
\end{align}
Here, $\mathbf{A}(r)$ is the linear mapping that transforms the amplitude $t(r)$ into the atomic orbital basis at geometry~$r$. Likewise, the left inverse  $\mathbf{A}^{-1}(r')$ transforms back into the molecular orbital basis at geometry $r'$. Combining both steps in the joint operator $\mathbf{L}(r',r)$ makes it possible to execute the transformation within memory scaling of $\mathcal{O}(V^2O^2)$ rather than $\mathcal{O}(N_b^4)$.
More explicitly, the transformed amplitude vector has single excitation components $\hat{t}^a_i(r',r):=\left[\mathbf{L}(r',r)t(r)\right]^a_i$ given by
\begin{align*}
    \hat{t}^a_i(r',r)= \sum_{bj} \ell^{\text{virt}}_{ab}(r',r) t^{b}_j(r)\ell^{\text{occ}}_{ij}(r',r),
\end{align*}
and double excitation components $\hat{t}^{ab}_{ij}(r',r):=\left[\mathbf{L}(r',r)t(r)\right]^{ab}_{ij}$, which are given by  
\begin{align*}
    \hat{t}^{ab}_{ij}(r',r)= \sum_{cdkl} \ell^{\text{virt}}_{ac}(r',r) \ell^{\text{virt}}_{bd}(r',r) t^{cd}_{kl}(r)\ell^{\text{occ}}_{ik}(r',r)\ell^{\text{occ}}_{jl}(r',r),
\end{align*}
where the matrices $\ell^{\text{virt}}(r',r)$ and $\ell^{\text{occ}}(r',r)$ are built from the coefficient matrix in the canonical orbital basis and the overlap matrix in the form of
\begin{align}
    \ell^{\text{virt}}_{ab}(r',r)= \sum_{\sigma}(CS)^*_{a \sigma}(r')C_{\sigma b}(r),
\end{align}
and 
\begin{align}
    \ell^{\text{occ}}_{ij}(r',r)= \sum_{\nu }C^*_{j \nu}(r)(SC)_{\nu i}(r'),
\end{align}
respectively.

Note that, in general, assembling $CS$ and $SC$ is done by a simple matrix multiplication scaling at most as $\mathcal{O}(N_b^3)$ in time and $\mathcal{O}(N_b^2)$ in memory. It is easy to see, that after the cheap assembling of $(SC)(r')$ and $(CS)(r')$, by performing one contraction after another the time complexity scaling for assembling the, more costly, transformed double excitation amplitudes $(\hat{t}^{ab}_{ij}(r',r))_{ij}^{ab}$ behaves like $\mathcal{O}(V^3O^2)$. Therefore, generating the transformed amplitudes is much cheaper than a true CCSD calculation.

Notably, due to the $S$-orthonormality of the coefficient matrix $C$, if $r'=r$ one has $\mathbf{L}(r,r)=\mathrm{Id}$ and so $\mathbf{L}(r,r)t(r)=t(r)$. As shown in~\cite{beck2026regularityinterpolationcoupledcluster}, $\mathbf{A}(r)t(r)$ is not affected by the orbital energy crossings or the potential sign flips arising from ambiguities in the underlying Hartree–Fock calculations, and therefore exhibits smooth behavior. Using the previous observations, given two nuclear geometries $r'$ and $r$ one can write
\begin{align*}
    \|t(r')-\mathbf{L}(r',r)t(r)\|&=\|\mathbf{A}^{-1}(r')[\mathbf{A}(r')t(r')-\mathbf{A}(r)t(r)]\|\\
    &\le \|\mathbf{A}^{-1}(r')\|\|\mathbf{A}(r')t(r')-\mathbf{A}(r)t(r)\|.
\end{align*}
Hence, the difference of the amplitude $t(r')$ and the amplitude that was transformed or shifted from geometry $r$ to $r'$ via $\mathbf{L}$ is split into a part $\|\mathbf{A}^{-1}(r')\|$ that only depends on $r'$ and the difference of the amplitudes in atomic orbital space. Since, the amplitudes in the atomic orbital basis are proven to be smooth, see \cite{beck2026regularityinterpolationcoupledcluster},
\begin{align*}
    \|\mathbf{A}(r')t(r')-\mathbf{A}(r)t(r)\|\longrightarrow 0,
\end{align*}
as $\|r'-r\|\rightarrow 0$ in a continuous manner. Therefore, a much better and more reliable extrapolation ansatz than \eqref{extrap_approx}, is given by
\begin{align}\label{transformed_extrap_ansatz}
    t(r^*)\approx \tilde{t}=\sum^{q}_{j=1}\alpha_j \mathbf{L}(r^*,r_j)t(r_j).
\end{align}
Note that the same transformation and extrapolation ansatz can also be applied to the Lambda amplitudes.

Besides the improved extrapolation ansatz, the transformation $\mathbf{L}$ also gives access to new descriptors $d$. An intuitive descriptor for quantities on the coupled cluster level of theory are the MP2 amplitudes $t^{\text{MP2}}(r)$, but they suffer from the same orbital energy crossings, which occur at the Hartree–Fock level of theory. Therefore, leveraging the atomic orbital transformation, a new possible descriptor can be the smooth MP2 amplitudes in atomic orbital basis $d(r)=\mathbf{A}(r)t^{\text{MP2}}(r)$. The descriptor in this form is intractable, since in the atomic orbital basis the memory scaling again reaches $\mathcal{O}(N_b^4)$. This issue can be circumvented by choosing $d(r)=\mathbf{L}(r^*,r)t^{\text{MP2}}(r)$ and doing a modified least square fitting in the descriptor space
\begin{align}\label{MP2_fitting}
    \alpha^*:=\argmin_{\alpha_1,\ldots,\alpha_q}\big\|d(r^*)-\sum^{q}_{j=1}\alpha_j \mathbf{L}(r^*,r_j)d(r_j)\big\|^2,
\end{align}
with the usual constraint $\sum^{q}_{j=1}\alpha_j=1$. The transformed least square minimization effectively fits the MP2 amplitudes in atomic orbital basis, that are smooth in $r$, without ever entering the memory intensive atomic orbital space explicitely.

\section{Results}
\label{sec:results}
Geometry optimizations have been performed using geomeTRIC~\cite{wang2016geometry,park2025high} as the main driver, which in turns uses CFOUR as the electronic structure engine \cite{10.1063/5.0004837, CFOUR_software}, and a custom Python module for providing initial guess amplitudes, which implements the extrapolation algorithm and the orbital transformation described in the previous section.
The convergence criteria in geomeTRIC were set to $|\Delta E|$ < 1.0e-06 $E_h$, RMS-Grad < 1.0e-05 $E_h/a_0$, Max-Grad < 4.5e-04 $E_h/a_0 $, RMSD <1.2e-03 Å and Max-Disp <1.8e-03 Å, where RMSD is the root-mean-square displacement.
All systems analyzed numerically are initialized with the starting nuclear geometry, obtained via a DFT geometry optimization in PySCF, with the functional $\omega$B97X-V and the def2-TZVP basis set \cite{PYSCF, Pyscf_geom, 10.1039/b508541a, 10.1039/c3cp54374a}. The DFT level geometry optimizations have been accelerated using the DFT version of G-Ext \cite{Polack17102020, 10.1021/acs.jctc.1c00751, 10.1021/acs.jpclett.3c02098}.  For all the investigated molecules, CC calculations have been performed using the cc-pVTZ basis set\cite{10.1063/1.456153, 10.1063/1.464303, 10.1063/1.478678}.

\subsection{Necessity of the transformation in the extrapolation}

\begin{figure}[H]
    \centering
    \includegraphics[width=0.7\linewidth]{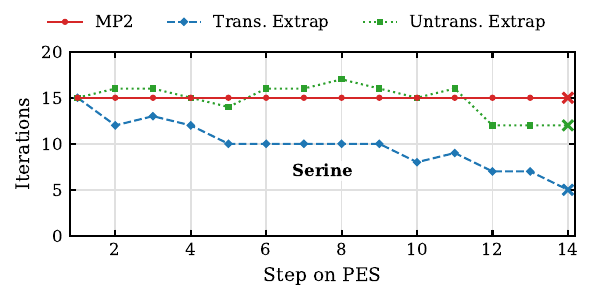}
    \caption{Number of necessary iterations of a coupled cluster calculation per step on the potential energy surface for the default MP2, extrapolation with transformed amplitudes and untransformed amplitudes.}
    \label{fig:trafo_no_trafo_comp}
\end{figure}

The necessity and success of the transformed extrapolation ansatz \eqref{transformed_extrap_ansatz} are depicted in Figure \ref{fig:trafo_no_trafo_comp}. For both the extrapolations —with and without the amplitude transformation— the choice of the parameter $\theta$ (see \eqref{extrap_parameter}) was with descriptor $d(r)=\mathbf{L}(r^*,r)t^{MP2}(r)$, considering $q=q_{\text{max}}$, i.e. all previous snapshots, and Tikhonov constant $\varepsilon=$1.0e-06.

Without the transformation, the extrapolation ansatz \eqref{extrap_approx} needs an average of $14.9$ iterations per step on the PES. 
This is only barely below the $15$-iteration average of the MP2 initial guess, and for some cases --- possibly in correspondence of significant orbital crossings --- the untransformed strategy performs even worse than the MP2 initial guess.
In contrast, the extrapolation ansatz with the transformation \eqref{transformed_extrap_ansatz} consistently outperforms the MP2 initial guess by a large margin and on average only needs $9.9$ iterations per step on the PES with a systematic improvement as the geometry optimizer converges.

\subsection{Extrapolation parameters} This section is devoted to finding a good choice of parameters $\theta=(d,q,\varepsilon)$ for the extrapolation, by doing a preliminary benchmark on a subset of the systems considered. Note that one cannot expect to find a $\theta$ that is optimal for every molecule. One reason is that the number of optimization steps required to reach convergence on the PES varies substantially across different molecules. For some molecules, the converged DFT geometry already provides a strong predictor, requiring only a few optimization steps, whereas for others, a much longer trajectory is necessary. Therefore, the main focus lies in finding $\theta$ that provides a better outcome the more steps on the PES had to be taken, which is proportional to the overall computational effort required. The relative performance of the extrapolation scheme \eqref{transformed_extrap_ansatz} with respect to a parameter $\theta$ over the course of a geometry optimization is assessed through the \textit{geometric mean relative absolute error} (GMRAE)
\begin{align}
    E^\theta_{GMRAE}:=\left(\prod^{N_{\text{steps}}}_{i=1}\frac{\|t(r_i)-\tilde{t}^\theta_i\|}{\|t(r_i)\|} \right)^{\frac{1}{N_{\text{steps}}}},
\end{align}
where $\tilde{t}_i^\theta$ denotes the extrapolant with respect to the design parameter $\theta$ at the $i$th step on the potential energy surface and $N_{\text{steps}}$ the number of steps necessary until convergence. The GMRAE is chosen over a normal average as it penalizes outliers less severely, thus giving a more faithful correspondence between the error in the amplitudes and the actual total number of iterations over the optimization process.

\begin{figure*}[t!]
    \centering
    \includegraphics[width=\textwidth]{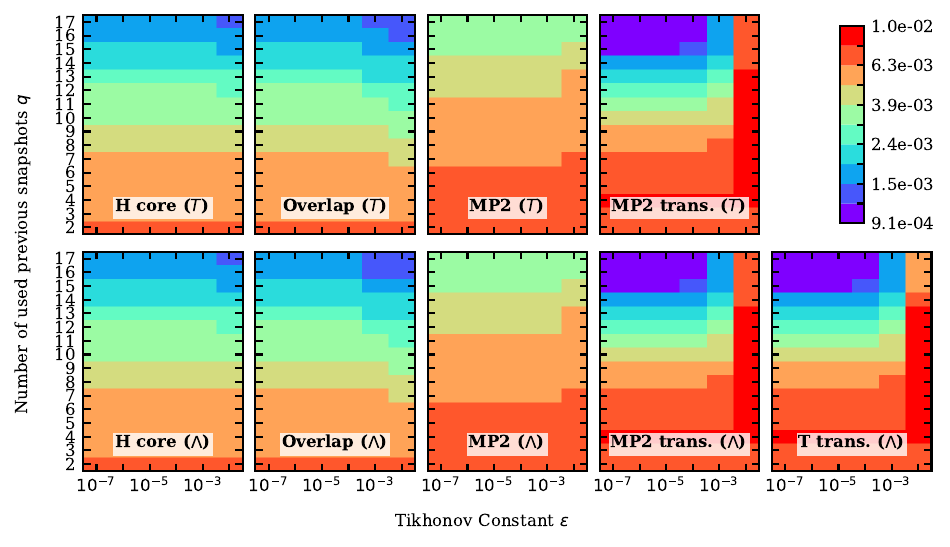}
    \caption{The GMRAE for the CC T-amplitudes (T) and $\Lambda$-amplitudes $(\Lambda)$ of alanine for various descriptors, numbers of previous snapshots $q$, and Tikhonov constants $\varepsilon$.}
    \label{fig:heatmaps}
\end{figure*}

 Since most of the molecules exhibit similar behavior, in Figure \ref{fig:heatmaps}, we illustrate, using alanine as an example, how $E_{GMRAE}^\theta$, for the CC- and $\Lambda$-amplitudes, varies across different choices of $\theta=(d,q,\varepsilon)$. Similar plots for other molecules are given in Figures S1--S7 in the supplementary material.
For the CC-amplitudes four different descriptors $d$ are considered. While the core Hamiltonian and overlap matrix are known descriptors that have previously been used in the context of DFT, we also consider two novel descriptors, the MP2 amplitudes and the transformed MP2 amplitudes following the scheme in \eqref{MP2_fitting}. For the $\Lambda$-amplitudes, a fifth descriptor is considered, namely the transformed converged CC-amplitudes. Figure \ref{fig:heatmaps} illustrates the qualitative behavior exhibited by most molecules considered in the investigation of the performance dependence on $\theta$.
First and foremost, the MP2 descriptor shows the worst relative error independent of the choice of $q$ and $\varepsilon $. This is to be expected, since the untransformed MP2 amplitudes are affected by index reorderings caused by orbital energy crossings, making a comparison with previously considered quantities notoriously difficult. Secondly, the core Hamiltonian and overlap descriptors benefit from a stronger Tikhonov regularization, particularly for larger values of $q$. This can be linked to the increasing likelihood of linear dependencies as more previous geometries are considered. For the transformed MP2 and transformed CC-amplitude descriptors, this phenomenon is not observed to the same extent. This can be attributed to two possible reasons. On the one hand, these descriptors have a significantly higher dimensionality than the core Hamiltonian and overlap, making the occurrence of linear dependencies less likely. On the other hand, the effect of the regularization depends on the norms of the descriptors themselves, which may differ substantially between the considered quantities.
For what concerns the $\Lambda$-amplitudes, we observe a similar picture with, in addition, the transformed CC T-amplitude descriptor showing nearly an identical performance as the transformed MP2 amplitudes descriptor.

All things considered, the final recommendation is to choose $\theta=(d,q,\varepsilon)$ such that $d$ is the transformed MP2 descriptor for both the CC T-amplitudes and the $\Lambda$-amplitudes, $q=q_{\text{max}}$, where $q_{\text{max}}$ denotes the largest value permitted by the available hardware under the memory requirement of $\mathcal{O}(q\,V^2O^2)$ in RAM, and the Tikhonov constant $\varepsilon=$1.0e-06.

\subsection{Comparing different initial guess strategies}

In most quantum chemistry packages, the MP2 amplitudes evaluated at the current geometry serve as the initial guess for the coupled cluster solver. While this strategy entirely circumvents the problem of orbital energy crossings, it does so at the expense of ignoring information from previous calculations. Two natural alternative extrapolation strategies would be directly using the previous amplitudes, or using the transformed previous amplitudes by taking advantage of the transformation $\mathbf{L}$, to counteract possible phenomena caused by orbital energy crossings.

\begin{table}[htbp]
    \centering
    \caption{Average number of iterations per step on the potential energy surface for the different initial guess methods.}
    \label{tab:avg_iterations}
    \small 
    
    \begin{tabular*}{\linewidth}{@{\extracolsep{\fill}} l c c c c}
        \toprule
        \textbf{Molecule} & \textbf{MP2} & \textbf{Prev.$^a$} & \textbf{Trans.$^b$} & \textbf{Extrap.} \\
        \midrule
        Pyrrole      & 15.0 & 15.2 & 12.3 & 10.6 \\
        Toluene      & 14.0 & 12.6 & 8.6 &  7.9 \\
        Benzaldehyde & 16.0 & 17.9 & 11.9 & 11.3 \\
        Cytosine     & 18.0 & 14.3 & 11.3 &  9.5 \\
        Uracil       & 17.0 & 15.4 & 10.3 &  9.9 \\
        Isobutane    & 12.0 & 12.7 & 11.8 &  9.5 \\
        Serine       & 15.0 & 13.3 & 11.0 &  9.9 \\
        Ethyl Acetate    & 15.0 & 15.6 & 10.6 &  9.8 \\
        \bottomrule
        \multicolumn{5}{l}{\footnotesize $^a$ Previous Amplitudes, $^b$ Transformed Prev. Amplitudes}
    \end{tabular*}
\end{table}

Table \ref{tab:avg_iterations} compares, for a collection of molecules, the extrapolation presented in the Theory section, with the three strategies outlined above, namely the MP2 default initial guess, the untransformed CC T-amplitudes at the previous step and the transformed CC T-amplitudes at the previous step.
For the extrapolation method, we used the extrapolation parameters determined in the previous section, i.e. $\theta=(d,q,\varepsilon)$ with $d$ the transformed MP2 descriptor, $q=q_{\text{max}}$ and $\varepsilon=$1.0e-06.

For all molecules considered in Table \ref{tab:avg_iterations},  the RMSD of the converged nuclear coordinates across all initial guess strategies varied on average only on the order of 1.0e-06 Å. 
The average number of iterations demonstrates a consistent advantage of the extrapolation method over all alternative initial guess strategies. Furthermore, it shows that using the previous geometry without the transformation is less reliable than employing the standard MP2-based initial guess.

\begin{figure}[H]
    \centering
    \includegraphics[width=0.7\textwidth]{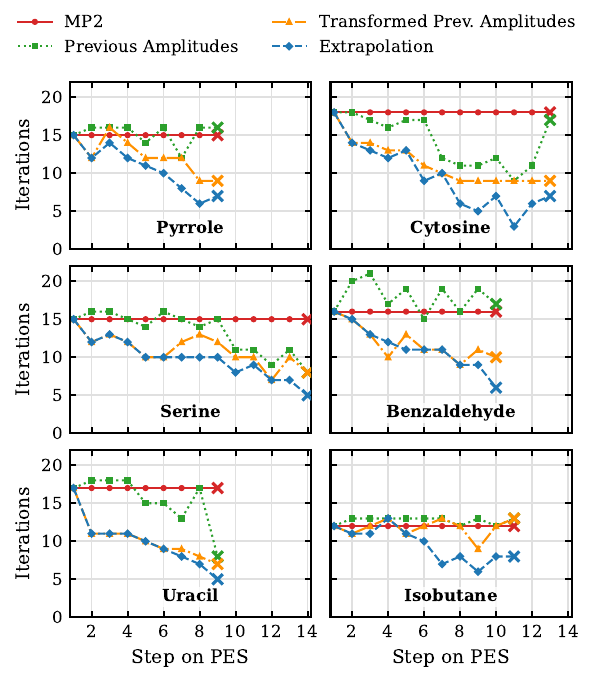}
    \caption{Comparison of the number of iterations per step on the potential energy surface for the different initial guess methods.}
    \label{fig:iteration_grid}
\end{figure}

Figure \ref{fig:iteration_grid} shows the number of iterations required at each step of a geometry optimization, thereby providing a qualitative comparison of the different approaches.
It is evident that the default MP2 approach is highly stable, whereas using the previous amplitudes proves unreliable and exhibits significant fluctuations. Transforming the amplitudes at the previous geometry provides a consistent, systematic improvement over the MP2 approach, likely because the transformation mitigates most of the effects caused by reorderings. Finally, by utilizing data from more than just the last geometry step, the extrapolation method achieves even fewer required iterations for almost all steps along the PES.

\subsection{Benchmarks}

CCSD geometry optimizations have been carried out for a collection of 32 molecules by using the cc-pVTZ basis set. The simulations have been executed both with the extrapolation strategy introduced in the present work and the default strategy of using the MP2 initial guess for the coupled cluster solver and the converged CCSD amplitudes as an initial guess for the Lambda solver. The starting geometries were chosen according to the description at the beginning of Section Results, i.e. from converged DFT geometry optimization.

\begin{figure*}[t!]
    \centering
    \includegraphics[width=\textwidth]{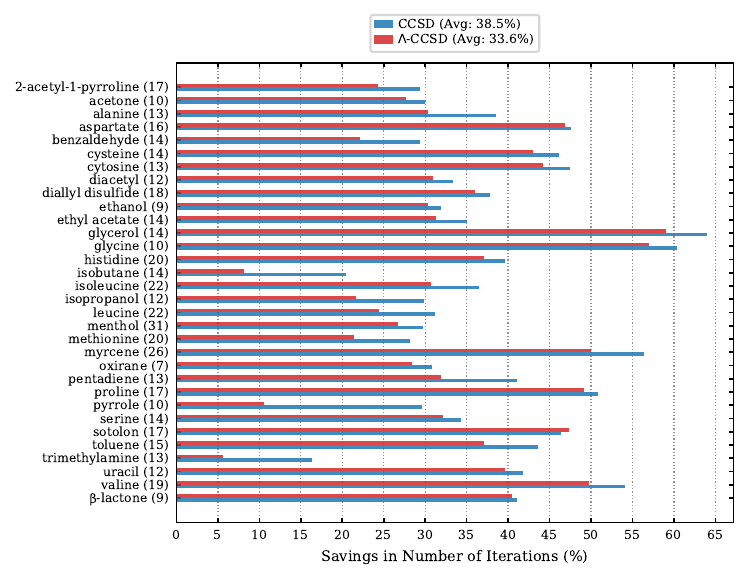}
    \caption{Number of Iterations saved when using the extrapolation initial guess compared to the MP2 amplitudes CCSD initial guess and the converged CCSD amplitudes Lambda initial guess. The numbers next to the molecules name represent its respective number of nuclei.}
    \label{fig:benchmarks}
\end{figure*}

Figure \ref{fig:benchmarks} shows the savings in the number of necessary iterations with the parameter $\theta=(d,q,\varepsilon)$ given by the transformed MP2 amplitude descriptor $d(r)=\mathbf{L}(r^*,r)t^{MP2}(r)$, number of previous snapshots $q=25$ and Tikhonov regularization parameter $\varepsilon=$1.0e-06, for both the CCSD Newton method and the Lambda Newton method respectively. The value $q=25$ was chosen to enable the investigation of larger molecules. Note, as indicated in Figure \ref{fig:heatmaps}, using $q>25$ should lead to even better results.

We observe reductions of the total number of CCSD iterations up to 64\% and Lambda iterations up to 59\%, while averaging across all molecules considered at 38.5\% and 33.6\% respectively. For all molecules considered the number of necessary steps on the PES until convergence was the same with and without the extrapolation strategy, and the average RMSD of the final geometries was 1.435e-06 Å, with the worst case being isopropanol at 2.072e-05 Å. A more detailed comparison for the geometries and the energies during the geometry optimizations is given in the Table S1 in the supplementary material.

\FloatBarrier

\section{Conclusions}\label{conclusion}

In the present work, we introduced an extrapolation strategy for geometry optimization at the coupled cluster level of theory. The method uses a transformation that makes the coupled cluster and Lambda amplitudes comparable to those from previous geometry steps, thereby making it possible to build sophisticated initial guesses for the quasi-Newton solver at new geometries. Building the extrapolants scales as $\mathcal{O}(V^3O^2)$ in time and $\mathcal{O}(V^2 O^2)$ in memory, thus introducing negligible computational overhead to the coupled cluster calculation itself.

To obtain an appropriate set of extrapolation parameters $\theta=(d,q,\varepsilon)$, a heuristic study was conducted, showing that the best performance are achieved by using the transformed MP2 amplitudes as a descriptor $d$, as many previous snapshots $q$ as feasible given the available hardware, and a small regularization constant $\varepsilon =$1.0e-06. Using the determined extrapolation parameters, a study on a small set of molecules was conducted, comparing the number of iterations required for the overall geometry optimizations when using our extrapolation method to the same number when using the standard MP2 initial guess strategy, as well as two more guesses, namely, amplitudes and transformed amplitudes from the previous optimization point. 

Benchmark simulations on a collection of over 30 different molecules showed that, on average the initial guesses provided by the extrapolation method reduce the number of necessary CC Newton iterations by 38.5\% and Lambda Newton iterations by 33.6\% with respect to the standard MP2 guess over the course of a complete geometry optimization. The converged nuclear geometries from the benchmark experiments remained well within the geomeTRIC optimizer's RMSD convergence threshold of 1.2e-03 Å, showing an average RMSD of 1.435e-06 Å (2.072e-05 Å at worst) compared to the reference calculations without the extrapolation strategy. Similarly, compared to the reference calculation without extrapolation, the average energy difference was 2.549e-6 kcal/mol, with a maximal difference of 1.255e-05 kcal/mol. In other words, the extrapolation scheme proposed in this contribution achieves sizeable reductions in computational effort without affecting the accuracy of the results. 

This work shows that extrapolation techniques can be successfully used to speed up Coupled Cluster calculations. The application of such techniques to CC theory comes with different challenges than the ones encountered in density functional theory, as the dependence of the CC amplitudes on the Hartree-Fock orbitals can lead to discontinuities whenever two orbitals are swapped along the potential energy surface. These difficulties were overcome by using an appropriate transformation that renders the transformed amplitudes a smooth function of the molecular geometry. 

In this contribution, we have explored geometry optimizations at the CCSD level of theory. An extension to different Coupled Cluster approximations is certainly possible, but not completely straightforward, as transforming higher order amplitudes (e.g., triples) can become computationally expensive. The present machinery can also be extended to saddle point optimizations, which are of interest in reactivity studies, and in principle even to ab-initio molecular dynamics.

\section*{Acknowledgements}

M.N. and B.S. acknowledge the Deutsche Forschungsgemeinschaft (DFG, German Research Foundation) for supporting this work by funding EXC2075 – 390740016 under Germany’s Excellence Strategy. We acknowledge the support from the Stuttgart Center for Simulation Science (SimTech).

\section*{Supplementary material}

In the supplementary material the extrapolation parameter experiments for additional molecules are presented. Additionally it provides a table of all the RMSDs and energy errors at the final geometries, as well as their averages along the entire optimization trajectory of the benchmark experiments.


\printbibliography

@article{Polack17102020,
author = {É. Polack and A. Mikhalev and G. Dusson and B. Stamm and F. Lipparini},
title = {An approximation strategy to compute accurate initial density matrices for repeated self-consistent field calculations at different geometries},
journal = {Molecular Physics},
volume = {118},
number = {19-20},
pages = {e1779834},
year = {2020},
publisher = {Taylor \& Francis},
doi = {10.1080/00268976.2020.1779834},


URL = { 
    
        https://doi.org/10.1080/00268976.2020.1779834
    
    

},
eprint = { 
    
        https://doi.org/10.1080/00268976.2020.1779834
    
    

}

}

@article{10.1021/acs.jctc.4c01417,
    author = {Askarpour, Zahra and Nottoli, Michele and Stamm, Benjamin},
    title = {Grassmann Extrapolation
for Accelerating Geometry
Optimization},
    journal = {Journal of Chemical Theory and Computation},
    volume = {21},
    number = {4},
    pages = {1643-1651},
    year = {2025},
    month = {02},
    issn = {1549-9618},
    doi = {10.1021/acs.jctc.4c01417},
    url = {https://doi.org/10.1021/acs.jctc.4c01417},
    eprint = {https://pubs.acs.org/jctcce/article-pdf/21/4/1643/41250390/ct4c01417.pdf},
}

@article{10.1021/acs.jctc.1c00751,
    author = {Polack, Étienne and Dusson, Geneviève and Stamm, Benjamin and Lipparini, Filippo},
    title = {Grassmann Extrapolation of Density Matrices for Born–Oppenheimer
Molecular Dynamics},
    journal = {Journal of Chemical Theory and Computation},
    volume = {17},
    number = {11},
    pages = {6965-6973},
    year = {2021},
    month = {10},
    issn = {1549-9618},
    doi = {10.1021/acs.jctc.1c00751},
    url = {https://doi.org/10.1021/acs.jctc.1c00751},
    eprint = {https://pubs.acs.org/jctcce/article-pdf/17/11/6965/13174658/ct1c00751.pdf},
}

@article{10.1021/acs.jpclett.3c02098,
    author = {Pes, Federica and Polack, Étienne and Mazzeo, Patrizia and Dusson, Geneviève and Stamm, Benjamin and Lipparini, Filippo},
    title = {A Quasi Time-Reversible
Scheme Based on Density Matrix
Extrapolation on the Grassmann Manifold for Born–Oppenheimer
Molecular Dynamics},
    journal = {The Journal of Physical Chemistry Letters},
    volume = {14},
    number = {43},
    pages = {9720-9726},
    year = {2023},
    month = {10},
    issn = {1948-7185},
    doi = {10.1021/acs.jpclett.3c02098},
    url = {https://doi.org/10.1021/acs.jpclett.3c02098},
    eprint = {https://pubs.acs.org/jpclcd/article-pdf/14/43/9720/3385263/jz3c02098.pdf},
}

@article{10.1063/5.0323068,
    author = {Nottoli, Michele and Castrignano', Damiano and Mazzeo, Patrizia and Cupellini, Lorenzo and Stamm, Benjamin},
    title = {Making excited state MD faster: Extrapolation of transition densities for TD-DFT calculations},
    journal = {The Journal of Chemical Physics},
    volume = {164},
    number = {15},
    pages = {154103},
    year = {2026},
    month = {04},
    issn = {0021-9606},
    doi = {10.1063/5.0323068},
    url = {https://doi.org/10.1063/5.0323068},
    eprint = {https://pubs.aip.org/aip/jcp/article-pdf/doi/10.1063/5.0323068/20975029/154103_1_5.0323068.pdf},
}

@article{10.1063/5.0141145,
    author = {Schrader, Simon Elias and Kvaal, Simen},
    title = {Accelerated coupled cluster calculations with Procrustes orbital interpolation},
    journal = {The Journal of Chemical Physics},
    volume = {158},
    number = {11},
    pages = {114116},
    year = {2023},
    month = {03},
    issn = {0021-9606},
    doi = {10.1063/5.0141145},
    url = {https://doi.org/10.1063/5.0141145},
    eprint = {https://pubs.aip.org/aip/jcp/article-pdf/doi/10.1063/5.0141145/16792499/114116_1_online.pdf},
}

@misc{beck2026regularityinterpolationcoupledcluster,
      title={On the Regularity and Interpolation of Coupled Cluster Amplitudes in Canonical Orbital Basis}, 
      author={Jonas Beck and Benjamin Stamm},
      year={2026},
      eprint={2605.22584},
      archivePrefix={arXiv},
      primaryClass={math.NA},
      url={https://arxiv.org/abs/2605.22584}, 
}

@book{shavitt2009many,
  title={Many-Body Methods in Chemistry and Physics: MBPT and Coupled-Cluster Theory},
  author={Shavitt, I. and Bartlett, R.J.},
  isbn={9780521818322},
  lccn={2009517972},
  series={Cambridge Molecular Science},
  url={https://books.google.de/books?id=SWw6ac1NHZYC},
  year={2009},
  publisher={Cambridge University Press}
}

@article{10.1063/1.443164,
    author = {Purvis, George D., III and Bartlett, Rodney J.},
    title = {A full coupled‐cluster singles and doubles model: The inclusion of disconnected triples},
    journal = {The Journal of Chemical Physics},
    volume = {76},
    number = {4},
    pages = {1910-1918},
    year = {1982},
    month = {02},
    issn = {0021-9606},
    doi = {10.1063/1.443164},
    url = {https://doi.org/10.1063/1.443164},
    eprint = {https://pubs.aip.org/aip/jcp/article-pdf/76/4/1910/18935003/1910_1_online.pdf},
}

@book{1212166, author = {Quarteroni, Alfio and Sacco, Riccardo and Saleri, Fausto}, title = {Numerical Mathematics (Texts in Applied Mathematics)}, year = {2006}, isbn = {3540346589}, publisher = {Springer-Verlag}, address = {Berlin, Heidelberg} }

@article{COESTER1960477,
title = {Short-range correlations in nuclear wave functions},
journal = {Nuclear Physics},
volume = {17},
pages = {477-485},
year = {1960},
issn = {0029-5582},
doi = {https://doi.org/10.1016/0029-5582(60)90140-1},
url = {https://www.sciencedirect.com/science/article/pii/0029558260901401},
author = {F. Coester and H. Kümmel}
}

@article{Schneider2009,
author = {Schneider, Reinhold},
year = {2009},
month = {08},
pages = {433-471},
title = {{Analysis of the projected coupled cluster method in electronic structure calculation}},
volume = {113},
journal = {Numerische Mathematik},
doi = {10.1007/s00211-009-0237-3}
}

@article{ rohwedder1,
	author = {{Rohwedder, Thorsten}},
	title = {The continuous Coupled Cluster formulation for the electronic
          Schrödinger equation},
	DOI= "10.1051/m2an/2012035",
	url= "https://doi.org/10.1051/m2an/2012035",
	journal = {ESAIM: M2AN},
	year = 2013,
	volume = 47,
	number = 2,
	pages = "421-447",
	month = "",
}

@book{helgakar,
  address = {Chichester},
  author = {Helgaker, T and J{\o}rgensen, P and Olsen, J},
  publisher = {John Wiley \& Sons, LTD},
  title = {{Molecular Electronic Structure Theory}},
  year = 2000
}

@article{10.1063/5.0004837,
    author = {Matthews, Devin A. and Cheng, Lan and Harding, Michael E. and Lipparini, Filippo and Stopkowicz, Stella and Jagau, Thomas-C. and Szalay, Péter G. and Gauss, Jürgen and Stanton, John F.},
    title = {Coupled-cluster techniques for computational chemistry: The CFOUR program package},
    journal = {The Journal of Chemical Physics},
    volume = {152},
    number = {21},
    pages = {214108},
    year = {2020},
    month = {06},
    issn = {0021-9606},
    doi = {10.1063/5.0004837},
    url = {https://doi.org/10.1063/5.0004837},
    eprint = {https://pubs.aip.org/aip/jcp/article-pdf/doi/10.1063/5.0004837/16669743/214108_1_online.pdf},
}

@misc{CFOUR_software,
  author       = {Stanton, J. F. and Gauss, J. and Cheng, L. and Harding, M. E. and Matthews, D. A. and Szalay, P. G. and Stopkowicz, S. and Lipparini, F. and Asthana, A. and Auer, A. A. and others},
  title        = {{CFOUR}, a quantum chemical program package},
  note         = {with contributions from and the integral packages MOLECULE (J. Almlöf and P. R. Taylor), PROPS (P. R. Taylor), ABACUS (T. Helgaker, H. J. Aa. Jensen, P. Jørgensen, and J. Olsen), and ECP routines by A. V. Mitin and C. van Wüllen. For the current version, http://www.cfour.de.}
}

@article{PYSCF,
author = {Sun, Qiming and Berkelbach, Timothy C. and Blunt, Nick S. and Booth, George H. and Guo, Sheng and Li, Zhendong and Liu, Junzi and McClain, James D. and Sayfutyarova, Elvira R. and Sharma, Sandeep and Wouters, Sebastian and Chan, Garnet Kin-Lic},
title = {{PySCF: the Python-based simulations of chemistry framework}},
journal = {WIREs Computational Molecular Science},
volume = {8},
number = {1},
pages = {e1340},
doi = {https://doi.org/10.1002/wcms.1340},
url = {https://wires.onlinelibrary.wiley.com/doi/abs/10.1002/wcms.1340},
eprint = {https://wires.onlinelibrary.wiley.com/doi/pdf/10.1002/wcms.1340},
year = {2018}
}

@article{Pyscf_geom,
    author = {Wang, Lee-Ping and Song, Chenchen},
    title = {{Geometry optimization made simple with translation and rotation coordinates}},
    journal = {The Journal of Chemical Physics},
    volume = {144},
    number = {21},
    pages = {214108},
    year = {2016},
    month = {06},
    issn = {0021-9606},
    doi = {10.1063/1.4952956},
    url = {https://doi.org/10.1063/1.4952956},
    eprint = {https://pubs.aip.org/aip/jcp/article-pdf/doi/10.1063/1.4952956/15512057/214108\_1\_online.pdf},
}

@article{mller_plesset_1934, title={Note on an Approximation Treatment for Many-Electron Systems}, volume={46}, ISSN={0031-899X}, DOI={10.1103/PhysRev.46.618}, abstractNote={A perturbation theory is developed for treating a system of n electrons in which the Hartree-Fock solution appears as the zero-order approximation. It is shown by this development that the first order correction for the energy and the charge density of the system is zero. The expression for the second-order correction for the energy greatly simplifies because of the special property of the zero-order solution. It is pointed out that the development of the higher approximation involves only calculations based on a definite one-body problem.}, number={7}, journal={Physical Review}, publisher={Physical Review}, author={Møller, Chr. and Plesset, M. S.}, year={1934}, month=oct, pages={618–622} }

@article{doi:10.1137/1021044,
author = {Willoughby, Ralph A.},
title = {Solutions of Ill-Posed Problems (A. N. Tikhonov and V. Y. Arsenin)},
journal = {SIAM Review},
volume = {21},
number = {2},
pages = {266-267},
year = {1979},
doi = {10.1137/1021044},

URL = { 
    
        https://doi.org/10.1137/1021044
    
    

},
eprint = { 
    
        https://doi.org/10.1137/1021044
    
    

}

}

@article{10.1063/1.456153,
    author = {Dunning, Thom H., Jr.},
    title = {Gaussian basis sets for use in correlated molecular calculations. I. The atoms boron through neon and hydrogen},
    journal = {The Journal of Chemical Physics},
    volume = {90},
    number = {2},
    pages = {1007-1023},
    year = {1989},
    month = {01},
    issn = {0021-9606},
    doi = {10.1063/1.456153},
    url = {https://doi.org/10.1063/1.456153},
    eprint = {https://pubs.aip.org/aip/jcp/article-pdf/90/2/1007/18974738/1007_1_online.pdf},
}

@article{10.1063/1.464303,
    author = {Woon, David E. and Dunning, Thom H., Jr.},
    title = {Gaussian basis sets for use in correlated molecular calculations. III. The atoms aluminum through argon},
    journal = {The Journal of Chemical Physics},
    volume = {98},
    number = {2},
    pages = {1358-1371},
    year = {1993},
    month = {01},
    issn = {0021-9606},
    doi = {10.1063/1.464303},
    url = {https://doi.org/10.1063/1.464303},
    eprint = {https://pubs.aip.org/aip/jcp/article-pdf/98/2/1358/19124934/1358_1_online.pdf},
}

@article{10.1063/1.478678,
    author = {Wilson, Angela K. and Woon, David E. and Peterson, Kirk A. and Dunning, Thom H., Jr.},
    title = {Gaussian basis sets for use in correlated molecular calculations. IX. The atoms gallium through krypton},
    journal = {The Journal of Chemical Physics},
    volume = {110},
    number = {16},
    pages = {7667-7676},
    year = {1999},
    month = {04},
    issn = {0021-9606},
    doi = {10.1063/1.478678},
    url = {https://doi.org/10.1063/1.478678},
    eprint = {https://pubs.aip.org/aip/jcp/article-pdf/110/16/7667/19328616/7667_1_online.pdf},
}

@article{10.1039/b508541a,
    author = {Weigend, Florian and Ahlrichs, Reinhart},
    title = {Balanced basis sets of split valence, triple zeta valence and quadruple zeta valence quality for H to Rn: Design and assessment of accuracy},
    journal = {Physical Chemistry Chemical Physics},
    volume = {7},
    number = {18},
    pages = {3297-3305},
    year = {2005},
    month = {01},
    issn = {1463-9076},
    doi = {10.1039/b508541a},
    url = {https://doi.org/10.1039/b508541a},
    eprint = {https://pubs.rsc.org/cp/article-pdf/7/18/3297/182773/b508541a.pdf},
}

@article{10.1039/c3cp54374a,
    author = {Mardirossian, Narbe and Head-Gordon, Martin},
    title = {{$\omega$B97X-V}: A 10-parameter, range-separated hybrid, generalized gradient approximation density functional with nonlocal correlation, designed by a survival-of-the-fittest strategy},
    journal = {Physical Chemistry Chemical Physics},
    volume = {16},
    number = {21},
    pages = {9904-9924},
    year = {2014},
    month = {06},
    issn = {1463-9076},
    doi = {10.1039/c3cp54374a},
    url = {https://doi.org/10.1039/c3cp54374a},
    eprint = {https://pubs.rsc.org/cp/article-pdf/16/21/9904/3034317/c3cp54374a.pdf},
}

@article{https://doi.org/10.1002/jcc.70473,
author = {Spura, Thomas and Elgabarty, Hossam and Kühne, Thomas D.},
title = {Accelerated “On-the-Fly” Coupled-Cluster Path-Integral Molecular Dynamics: Impact of Nuclear Quantum Effects on An Asymmetric Proton},
journal = {Journal of Computational Chemistry},
volume = {47},
number = {21},
pages = {e70473},
doi = {https://doi.org/10.1002/jcc.70473},
url = {https://onlinelibrary.wiley.com/doi/abs/10.1002/jcc.70473},
eprint = {https://onlinelibrary.wiley.com/doi/pdf/10.1002/jcc.70473},
year = {2026}
}

@article{481c956f-fc76-3a47-912c-132550d4ccb6,
 ISSN = {00401706},
 URL = {http://www.jstor.org/stable/1267352},
 author = {Arthur E. Hoerl and Robert W. Kennard},
 journal = {Technometrics},
 number = {1},
 pages = {69--82},
 publisher = {{Taylor \& Francis, Ltd., American Statistical Association, American Society for Quality}},
 title = {Ridge Regression: Applications to Nonorthogonal Problems},
 urldate = {2026-09-15},
 volume = {12},
 year = {1970}
}

@article{helgaker_properties,
	author = {Helgaker, Trygve and Coriani, Sonia and Jørgensen, Poul and Kristensen, Kasper and Olsen, Jeppe and Ruud, Kenneth},
	doi = {10.1021/cr2002239},
	eprint = { https://doi.org/10.1021/cr2002239 },
	journal = {Chem. Rev.},
	number = {1},
	pages = {543–631},
	title = {Recent Advances in Wave Function-Based Methods of Molecular-Property Calculations},
	url = {https://doi.org/10.1021/cr2002239},
	volume = {112},
	year = {2012}}

@Inbook{Pulay1977grad,
author="Pulay, P{\'e}ter",
editor="Schaefer, Henry F.",
title="Direct Use of the Gradient for Investigating Molecular Energy Surfaces",
bookTitle="Applications of Electronic Structure Theory",
year="1977",
publisher="Springer US",
address="Boston, MA",
pages="153--185",
isbn="978-1-4684-8541-7",
doi="10.1007/978-1-4684-8541-7_4",
url="https://doi.org/10.1007/978-1-4684-8541-7_4"
}

@article{schlegel2011geometry,
  title={Geometry optimization},
  author={Schlegel, H Bernhard},
  journal={WIREs: Comput. Mol. Sci.},
  volume={1},
  number={5},
  pages={790--809},
  year={2011},
  publisher={Wiley Online Library}
}

@article{adamowicz1984lambda,
author = {Adamowicz, L. and Laidig, W. D. and Bartlett, R. J.},
title = {Analytical gradients for the coupled-cluster method},
journal = {Int. J. Quantum Chem.},
volume = {26},
number = {S18},
pages = {245-254},
doi = {https://doi.org/10.1002/qua.560260825},
url = {https://onlinelibrary.wiley.com/doi/abs/10.1002/qua.560260825},
eprint = {https://onlinelibrary.wiley.com/doi/pdf/10.1002/qua.560260825},
year = {1984}
}

@article{scheiner1987analytic,
  title={Analytic evaluation of energy gradients for the single and double excitation coupled cluster (CCSD) wave function: Theory and application},
  author={Scheiner, Andrew C and Scuseria, Gustavo E and Rice, Julia E and Lee, Timothy J and Schaefer III, Henry F},
  journal={J. Chem. Phys.},
  volume={87},
  number={9},
  pages={5361--5373},
  year={1987},
  publisher={American Institute of Physics}
}

@article{salter1989analytic,
    author = {Salter, E. A. and Trucks, Gary W. and Bartlett, Rodney J.},
    title = {Analytic energy derivatives in many‐body methods. I. First derivatives},
    journal = {J. Chem. Phys.},
    volume = {90},
    number = {3},
    pages = {1752-1766},
    year = {1989},
    month = {02},
    issn = {0021-9606},
    doi = {10.1063/1.456069},
    url = {https://doi.org/10.1063/1.456069},
    eprint = {https://pubs.aip.org/aip/jcp/article-pdf/90/3/1752/18975501/1752_1_online.pdf}
}

@article{GAUSS1991207rohf,
title = {Analytic energy gradients for open-shell coupled-cluster singles and doubles (CCSD) calculations using restricted open-shell Hartree—Fock (ROHF) reference functions},
journal = {Chem. Phys. Lett.},
volume = {182},
number = {3},
pages = {207-215},
year = {1991},
issn = {0009-2614},
doi = {https://doi.org/10.1016/0009-2614(91)80203-A},
url = {https://www.sciencedirect.com/science/article/pii/000926149180203A},
author = {Jürgen Gauss and Walter J. Lauderdale and John F. Stanton and John D. Watts and Rodney J. Bartlett}
}

@article{gauss1991qrhf,
    author = {Gauss, Jürgen and Stanton, John F. and Bartlett, Rodney J.},
    title = {Analytic evaluation of energy gradients at the coupled‐cluster singles and doubles level using quasi‐restricted Hartree–Fock open‐shell reference functions},
    journal = {J. Chem. Phys.},
    volume = {95},
    number = {4},
    pages = {2639-2645},
    year = {1991},
    month = {08},
    issn = {0021-9606},
    doi = {10.1063/1.460916},
    url = {https://doi.org/10.1063/1.460916},
    eprint = {https://pubs.aip.org/aip/jcp/article-pdf/95/4/2639/18993530/2639_1_online.pdf}
}

@article{scuseria1991analytic,
    author = {Scuseria, Gustavo E.},
    title = {Analytic evaluation of energy gradients for the singles and doubles coupled cluster method including perturbative triple excitations: Theory and applications to FOOF and Cr2},
    journal = {J. Chem. Phys.},
    volume = {94},
    number = {1},
    pages = {442-447},
    year = {1991},
    month = {01},
    issn = {0021-9606},
    doi = {10.1063/1.460359},
    url = {https://doi.org/10.1063/1.460359},
    eprint = {https://pubs.aip.org/aip/jcp/article-pdf/94/1/442/18990583/442_1_online.pdf}
}

@article{lee1991analytic,
    author = {Lee, Timothy J. and Rendell, Alistair P.},
    title = {Analytic gradients for coupled‐cluster energies that include noniterative connected triple excitations: Application to cis‐ and trans‐HONO},
    journal = {J. Chem. Phys.},
    volume = {94},
    number = {9},
    pages = {6229-6236},
    year = {1991},
    month = {05},
    issn = {0021-9606},
    doi = {10.1063/1.460411},
    url = {https://doi.org/10.1063/1.460411},
    eprint = {https://pubs.aip.org/aip/jcp/article-pdf/94/9/6229/18992725/6229_1_online.pdf}
}

@article{WATTS19921open,
title = {Open-shell analytical energy gradients for triple excitation many-body, coupled-cluster methods: MBPT(4), CCSD+T(CCSD), CCSD(T),and QCISD(T)},
journal = {Chem. Phys. Lett.},
volume = {200},
number = {1},
pages = {1-7},
year = {1992},
issn = {0009-2614},
doi = {https://doi.org/10.1016/0009-2614(92)87036-O},
url = {https://www.sciencedirect.com/science/article/pii/000926149287036O},
author = {John D. Watts and Jürgen Gauss and Rodney J. Bartlett}
}

@ARTICLE{Watts93,
   author       = {John D. Watts and Jürgen Gauss and Rodney J. Bartlett},
   year         = {1993},
   title        = {Coupled‐cluster methods with noniterative triple excitations for restricted open‐shell {Hartree–Fock} and other general single determinant reference functions. {Energies} and analytical gradients},
   journal      = {J. Chem. Phys.},
   volume       = {98},
   pages        = {8718-8733},
   doi={10.1063/1.464480}
}

@article{gauss2002ccsdt,
    author = {Gauss, Jürgen and Stanton, John F.},
    title = {Analytic gradients for the coupled-cluster singles, doubles, and triples (CCSDT) model},
    journal = {J. Chem. Phys.},
    volume = {116},
    number = {5},
    pages = {1773-1782},
    year = {2002},
    month = {02},
    issn = {0021-9606},
    doi = {10.1063/1.1429244},
    url = {https://doi.org/10.1063/1.1429244},
    eprint = {https://pubs.aip.org/aip/jcp/article-pdf/116/5/1773/19315619/1773_1_online.pdf}
}

@article{kallay2003analytic,
  title={Analytic first derivatives for general coupled-cluster and configuration interaction models},
  author={K{\'a}llay, Mih{\'a}ly and Gauss, J{\"u}rgen and Szalay, P{\'e}ter G},
  journal={J. Chem. Phys.},
  volume={119},
  number={6},
  pages={2991--3004},
  year={2003},
  publisher={American Institute of Physics}
}

@article{koch1990lagrangian,
  title={Coupled cluster energy derivatives. Analytic Hessian for the closed-shell coupled cluster singles and doubles wave function: Theory and applications},
  author={Koch, Henrik and Jensen, Hans Jo/rgen Aa and Jørgensen, Poul and Helgaker, Trygve and Scuseria, Gustavo E and Schaefer III, Henry F},
  journal={J. Chem. Phys.},
  volume={92},
  number={8},
  pages={4924--4940},
  year={1990},
  publisher={American Institute of Physics}
}

@article{helgaker1982simple,
author = {Helgaker, Trygve Ulf},
title = {Simple derivation of the potential energy gradient for an arbitrary electronic wave function},
journal = {Int. J. Quantum Chem.},
volume = {21},
number = {5},
pages = {939-940},
doi = {https://doi.org/10.1002/qua.560210520},
url = {https://onlinelibrary.wiley.com/doi/abs/10.1002/qua.560210520},
eprint = {https://onlinelibrary.wiley.com/doi/pdf/10.1002/qua.560210520},
year = {1982}
}

@article{helgaker1988analytical,
  title={Analytical calculation of geometrical derivatives in molecular electronic structure theory},
  author={Helgaker, Trygve and J{\o}rgensen, Poul},
  journal={Adv. Quantum Chem.},
  volume={19},
  pages={183--245},
  year={1988},
  publisher={Elsevier}
}

@article{hald2003lagrangian,
    author = {Hald, Kasper and Halkier, Asger and Jørgensen, Poul and Coriani, Sonia and Hättig, Christof and Helgaker, Trygve},
    title = {A Lagrangian, integral-density direct formulation and implementation of the analytic CCSD and CCSD(T) gradients},
    journal = {J. Chem. Phys.},
    volume = {118},
    number = {7},
    pages = {2985-2998},
    year = {2003},
    month = {02},
    issn = {0021-9606},
    doi = {10.1063/1.1531106},
    url = {https://doi.org/10.1063/1.1531106},
    eprint = {https://pubs.aip.org/aip/jcp/article-pdf/118/7/2985/19321945/2985_1_online.pdf},
}

@article{gauss1991coupled,
  title={{Coupled-cluster open-shell analytic gradients: Implementation of the direct product decomposition approach in energy gradient calculations}},
  doi={10.1063/1.460915},
  author={Gauss, J{\"u}rgen and Stanton, John F and Bartlett, Rodney J},
  journal={J. Chem. Phys.},
  volume={95},
  number={4},
  pages={2623--2638},
  year={1991},
  publisher={American Institute of Physics}
}

@ARTICLE{Scuseria88,
   author       = {Gustavo E. Scuseria and Henry F. {Schaefer III}},
   year         = {1988},
   title        = {Analytic evaluation of energy gradients for the single, double and linearized triple excitation coupled-cluster {CCSDT-1} wavefunction: {Theory} and applications},
   journal      = {Chem. Phys. Lett.},
   volume       = {146},
   pages        = {23-31},
   doi={10.1016/0009-2614(88)85042-5},
}

@ARTICLE{Gauss00,
   author       = {J{\"u}rgen Gauss and John F. Stanton},
   year         = {2000},
   title        = {Analytic first and second derivatives for the {CCSDT-n} (n=1–3) models: a first step towards the efficient calculation of {CCSDT} properties},
   journal      = {Phys. Chem. Chem. Phys.},
   volume       = {2},
   pages        = {2047-2059},
   doi={10.1039/A909820H}
}

@article{harding2008parallel,
  title={Parallel calculation of CCSD and CCSD(T) analytic first and second derivatives},
  author={Harding, Michael E and Metzroth, Thorsten and Gauss, J{\"u}rgen and Auer, Alexander A},
  journal={J. Chem. Theory Comput.},
  volume={4},
  number={1},
  pages={64--74},
  year={2008},
  publisher={ACS Publications}
}

@article{hampel1996local,
  title={Local treatment of electron correlation in coupled cluster theory},
  doi={10.1063/1.471289},
  author={Hampel, Claudia and Werner, Hans-Joachim},
  journal={J. Chem. Phys.},
  volume={104},
  number={16},
  pages={6286--6297},
  year={1996},
  publisher={American Institute of Physics}
}

@article{werner2011efficient,
  title={An efficient local coupled cluster method for accurate thermochemistry of large systems},
  author={Werner, Hans-Joachim and Sch{\"u}tz, Martin},
  journal={J. Chem. Phys.},
  volume={135},
  number={14},
  year={2011},
  pages={144116},
  publisher={AIP Publishing}
}

@article{wei2009local,
    author = {Li, Wei and Piecuch, Piotr and Gour, Jeffrey R. and Li, Shuhua},
    title = {Local correlation calculations using standard and renormalized coupled-cluster approaches},
    journal = {J. Chem. Phys.},
    volume = {131},
    number = {11},
    pages = {114109},
    year = {2009},
    month = {09},
    doi = {10.1063/1.3218842},
}

@article{rolik2011general,
  title={A general-order local coupled-cluster method based on the cluster-in-molecule approach},
  author={Rolik, Zolt{\'a}n and K{\'a}llay, Mih{\'a}ly},
  journal={J. Chem. Phys.},
  volume={135},
  number={10},
  year={2011},
  pages={104111},
  publisher={AIP Publishing}
}

@article{riplinger2013dlpno,
    author = {Riplinger, Christoph and Neese, Frank},
    title = {An efficient and near linear scaling pair natural orbital based local coupled cluster method},
    journal = {J. Chem. Phys.},
    volume = {138},
    number = {3},
    pages = {034106},
    year = {2013},
    month = {01},
    doi = {10.1063/1.4773581},
}

@article{guo2018dlpno,
    author = {Guo, Yang and Riplinger, Christoph and Becker, Ute and Liakos, Dimitrios G. and Minenkov, Yury and Cavallo, Luigi and Neese, Frank},
    title = {Communication: An improved linear scaling perturbative triples correction for the domain based local pair-natural orbital based singles and doubles coupled cluster method [DLPNO-CCSD(T)]},
    journal = {J. Chem. Phys.},
    volume = {148},
    number = {1},
    pages = {011101},
    year = {2018},
    month = {01},
    doi = {10.1063/1.5011798},
}

@article{epifanovsky2013general,
    author = {Epifanovsky, Evgeny and Zuev, Dmitry and Feng, Xintian and Khistyaev, Kirill and Shao, Yihan and Krylov, Anna I.},
    title = "{General implementation of the resolution-of-the-identity and Cholesky representations of electron repulsion integrals within coupled-cluster and equation-of-motion methods: Theory and benchmarks}",
    journal = {J. Chem. Phys.},
    volume = {139},
    number = {13},
    year = {2013},
    pages={134105},
    month = {10}}

@article{Nottoli2023,
    author = {Nottoli, Tommaso and Gauss, Jürgen and Lipparini, Filippo},
    title = "{A novel coupled-cluster singles and doubles implementation that combines the exploitation of point-group symmetry and Cholesky decomposition of the two-electron integrals}",
    journal = {J. Chem. Phys.},
    volume = {159},
    number = {23},
    pages = {231101},
    year = {2023},
    month = {12},
    issn = {0021-9606},
    doi = {10.1063/5.0175956},
    url = {https://doi.org/10.1063/5.0175956},
}

@article{deprince2013accuracy,
  title={Accuracy and efficiency of coupled-cluster theory using density fitting/cholesky decomposition, frozen natural orbitals, and at 1-transformed hamiltonian},
  author={DePrince III, A Eugene and Sherrill, C David},
  journal={J. Chem. Theory Comput.},
  volume={9},
  number={6},
  pages={2687--2696},
  year={2013},
  publisher={ACS Publications}
}

@article{schuetz2001,
    author = {Schütz, Martin and Werner, Hans-Joachim},
    title = {{Low-order scaling local electron correlation methods. IV. Linear scaling local coupled-cluster (LCCSD)}},
    journal = {J. Chem. Phys.},
    volume = {114},
    number = {2},
    pages = {661-681},
    year = {2001},
    month = {01},
    doi = {10.1063/1.1330207},
}

@article{neese2009efficient,
  title={Efficient and accurate local approximations to coupled-electron pair approaches: An attempt to revive the pair natural orbital method},
  author={Neese, Frank and Wennmohs, Frank and Hansen, Andreas},
  journal={J. Chem. Phys.},
  volume={130},
  number={11},
  year={2009},
  pages={114108},
  publisher={AIP Publishing}
}

@article{neese2009efficient2,
  title={Efficient and accurate approximations to the local coupled cluster singles doubles method using a truncated pair natural orbital basis},
  author={Neese, Frank and Hansen, Andreas and Liakos, Dimitrios G},
  journal={J. Chem. Phys.},
  volume={131},
  number={6},
  year={2009},
  pages={064103},
  publisher={AIP Publishing}
}

@article{kjaergaard2017divide,
  title={The divide--expand--consolidate coupled cluster scheme},
  author={Kj{\ae}rgaard, Thomas and Baudin, Pablo and Bykov, Dmytro and Kristensen, Kasper and J{\o}rgensen, Poul},
  journal={Wiley Interdiscip. Rev.: Comput. Mol. Sci.},
  volume={7},
  number={6},
  pages={e1319},
  year={2017},
  publisher={Wiley Online Library}
}

@article{peng2017low,
  title={Low-rank factorization of electron integral tensors and its application in electronic structure theory},
  author={Peng, Bo and Kowalski, Karol},
  journal={Chem. Phys. Lett.},
  volume={672},
  pages={47--53},
  year={2017},
  publisher={Elsevier}
}

@article{ma2018explicitly,
  title={Explicitly correlated local coupled-cluster methods using pair natural orbitals},
  author={Ma, Qianli and Werner, Hans-Joachim},
  journal={Wiley Interdiscip. Rev.: Comput. Mol. Sci.},
  volume={8},
  number={6},
  pages={e1371},
  year={2018},
  publisher={Wiley Online Library}
}

@article{parrish2012hyper,
    author = {Parrish, Robert M. and Hohenstein, Edward G. and Martínez, Todd J. and Sherrill, C. David},
    title = {Tensor hypercontraction. II. Least-squares renormalization},
    journal = {J. Chem. Phys.},
    volume = {137},
    number = {22},
    pages = {224106},
    year = {2012},
    month = {12},
    issn = {0021-9606},
    doi = {10.1063/1.4768233},
    url = {https://doi.org/10.1063/1.4768233},
    eprint = {https://pubs.aip.org/aip/jcp/article-pdf/doi/10.1063/1.4768233/13999837/224106_1_online.pdf}
}

@article{parrish2019rank,
    author = {Parrish, Robert M. and Zhao, Yao and Hohenstein, Edward G. and Martínez, Todd J.},
    title = {Rank reduced coupled cluster theory. I. Ground state energies and wavefunctions},
    journal = {J. Chem. Phys.},
    volume = {150},
    number = {16},
    pages = {164118},
    year = {2019},
    month = {04},
    issn = {0021-9606},
    doi = {10.1063/1.5092505},
    url = {https://doi.org/10.1063/1.5092505},
    eprint = {https://pubs.aip.org/aip/jcp/article-pdf/doi/10.1063/1.5092505/9648754/164118_1_online.pdf}
}

@article{hohenstein2022rank,
  title={Rank-reduced coupled-cluster. III. Tensor hypercontraction of the doubles amplitudes},
  author={Hohenstein, Edward G and Fales, B Scott and Parrish, Robert M and Mart{\'\i}nez, Todd J},
  journal={J. Chem. Phys.},
  volume={156},
  pages={054102},
  number={5},
  year={2022},
  publisher={AIP Publishing}
}

@article{datta2016analytic,
  title={Analytic energy derivatives for the calculation of the first-order molecular properties using the domain-based local pair-natural orbital coupled-cluster theory},
  author={Datta, Dipayan and Kossmann, Simone and Neese, Frank},
  journal={J. Chem. Phys.},
  volume={145},
  number={11},
  year={2016},
  pages={114101},
  publisher={AIP Publishing}
}

@article{rauhut2001analytical,
  title={Analytical energy gradients for local coupled-cluster methods},
  author={Rauhut, Guntram and Werner, Hans-Joachim},
  journal={Phys. Chem. Chem. Phys.},
  volume={3},
  number={22},
  pages={4853--4862},
  year={2001},
  publisher={Royal Society of Chemistry}
}

@article{feng2019implementation,
    author = {Feng, Xintian and Epifanovsky, Evgeny and Gauss, Jürgen and Krylov, Anna I.},
    title = "{Implementation of analytic gradients for CCSD and EOM-CCSD using Cholesky decomposition of the electron-repulsion integrals and their derivatives: Theory and benchmarks}",
    journal = {J. Chem. Phys.},
    volume = {151},
    number = {1},
    year = {2019},
    pages={014110 },
    month = {07}}

@article{schnack2022efficient,
  title={Efficient implementation of molecular {CCSD} gradients with {Cholesky-decomposed} electron repulsion integrals},
  doi={10.1063/5.0087261},
  author={Schnack-Petersen, Anna Kristina and Koch, Henrik and Coriani, Sonia and Kj{\o}nstad, Eirik F},
  journal={J. Chem. Phys.},
  volume={156},
  number={24},
  pages={244111},
  year={2022},
  publisher={AIP Publishing LLC}
}

@article{bozkaya2016analytic,
    author = {Bozkaya, Uğur and Sherrill, C. David},
    title = {Analytic energy gradients for the coupled-cluster singles and doubles method with the density-fitting approximation},
    journal = {J. Chem. Phys.},
    volume = {144},
    number = {17},
    pages = {174103},
    year = {2016},
    month = {05},
    issn = {0021-9606},
    doi = {10.1063/1.4948318},
    url = {https://doi.org/10.1063/1.4948318},
    eprint = {https://pubs.aip.org/aip/jcp/article-pdf/doi/10.1063/1.4948318/13987412/174103_1_online.pdf}
}

@article{bozkaya2017triples,
    author = {Bozkaya, Uğur and Sherrill, C. David},
    title = {Analytic energy gradients for the coupled-cluster singles and doubles with perturbative triples method with the density-fitting approximation},
    journal = {J. Chem. Phys.},
    volume = {147},
    number = {4},
    pages = {044104},
    year = {2017},
    month = {07},
    issn = {0021-9606},
    doi = {10.1063/1.4994918},
    url = {https://doi.org/10.1063/1.4994918},
    eprint = {https://pubs.aip.org/aip/jcp/article-pdf/doi/10.1063/1.4994918/15529493/044104_1_online.pdf}
}

@article{melega2026novel,
  title={A novel implementation of CCSD analytic gradients using Cholesky decomposition of the two-electron integrals and Abelian point-group symmetry},
  author={Melega, Luca and Nottoli, Tommaso and Gauss, Jürgen and Lipparini, Filippo},
  journal={J. Phys. Chem. A},
  volume={130},
  number={10},
  pages={2097--2111},
  year={2026},
  publisher={ACS Publications}
}

@article{Pedersen2024cd,
author = {Pedersen, Thomas Bondo and Lehtola, Susi and Fdez. Galván, Ignacio and Lindh, Roland},
title = {The versatility of the Cholesky decomposition in electronic structure theory},
journal = {WIREs Comput. Mol. Sci.},
volume = {14},
number = {1},
pages = {e1692},
doi = {https://doi.org/10.1002/wcms.1692},
year = {2024}
}

@article{wang2016geometry,
  title={Geometry optimization made simple with translation and rotation coordinates},
  author={Wang, Lee-Ping and Song, Chenchen},
  journal={J. Chem. Phys.},
  volume={144},
  number={21},
  year={2016},
  publisher={AIP Publishing}
}

@article{park2025high,
  title={High-throughput approach for minimum energy pathway search using the nudged elastic band method with efficient data handling and parallel computing},
  author={Park, Heejune and Pritchard, Benjamin P and Wang, Lee-Ping},
  journal={J. Chem. Theory Comput.},
  volume={21},
  number={23},
  pages={12048--12063},
  year={2025},
  publisher={ACS Publications}
}

@book{Fletcher1999,
address = {New York},
author = {Fletcher, Roger},
chapter = {5.2},
edition = {Second},
pages = {100--107},
publisher = {Wiley},
title = {{Practical Methods of Optimization}},
year = {1999}
}

\end{document}


\maketitle

\FloatBarrier

\begin{figure*}
    \centering
    \includegraphics[width=\textwidth]{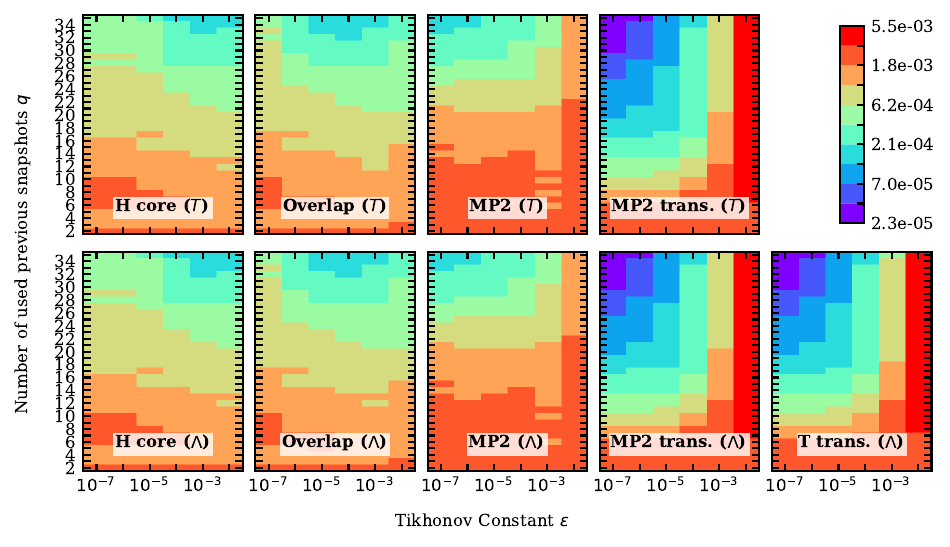}
    \caption{The GMRAE for the CC-amplitudes (T) and $\Lambda$-amplitudes $(\Lambda)$ of \textbf{glycine} for various descriptors, numbers of previous snapshots $q$, and Tikhonov constants $\varepsilon$.}
    \label{fig:heatmaps}
\end{figure*}

\begin{figure*}[t!]
    \centering
    \includegraphics[width=\textwidth]{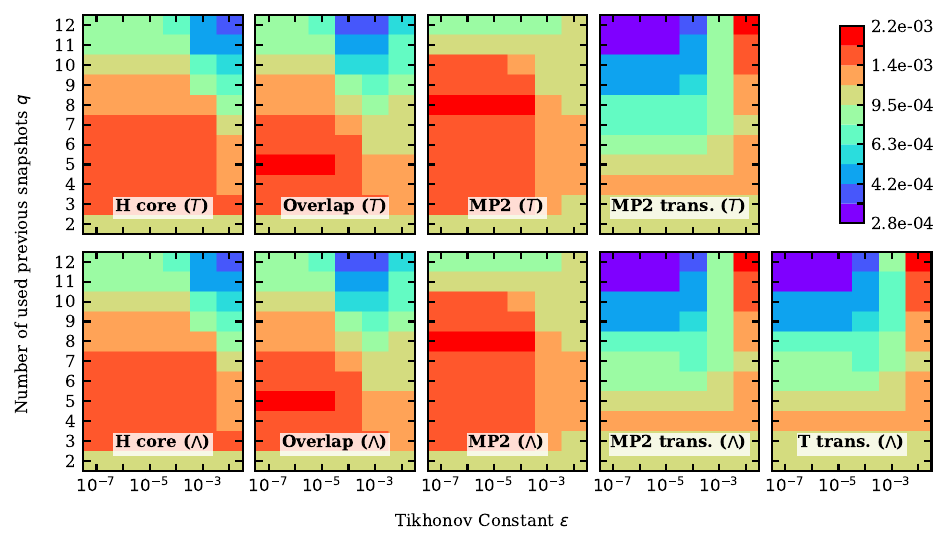}
    \caption{The GMRAE for the CC-amplitudes (T) and $\Lambda$-amplitudes $(\Lambda)$ of \textbf{cytosine} for various descriptors, numbers of previous snapshots $q$, and Tikhonov constants $\varepsilon$.}
    \label{fig:heatmaps}
\end{figure*}

\begin{figure*}[t!]
    \centering
    \includegraphics[width=\textwidth]{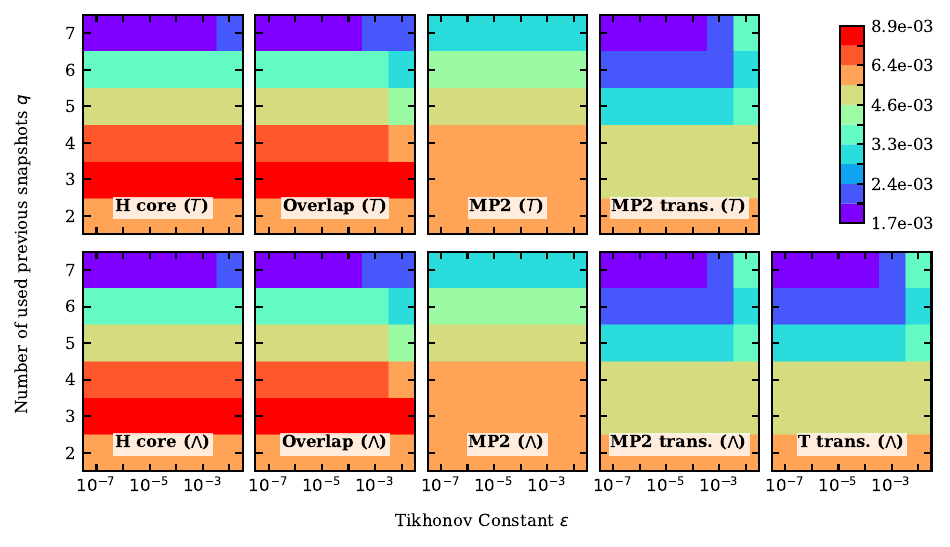}
    \caption{The GMRAE for the CC-amplitudes (T) and $\Lambda$-amplitudes $(\Lambda)$ of \textbf{ethyl acetate} for various descriptors, numbers of previous snapshots $q$, and Tikhonov constants $\varepsilon$.}
    \label{fig:heatmaps}
\end{figure*}

\begin{figure*}[t!]
    \centering
    \includegraphics[width=\textwidth]{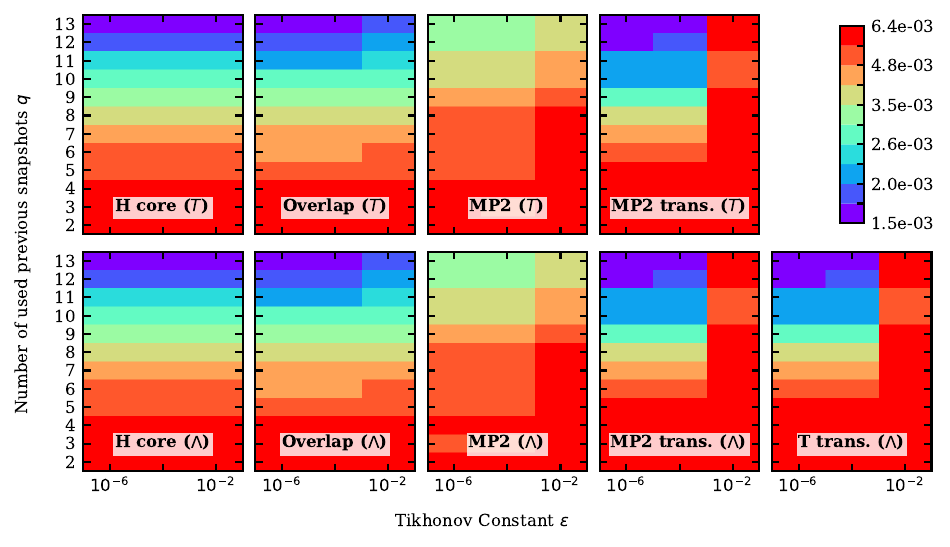}
    \caption{The GMRAE for the CC-amplitudes (T) and $\Lambda$-amplitudes $(\Lambda)$ of \textbf{serine} for various descriptors, numbers of previous snapshots $q$, and Tikhonov constants $\varepsilon$.}
    \label{fig:heatmaps}
\end{figure*}

\begin{figure*}[t!]
    \centering
    \includegraphics[width=\textwidth]{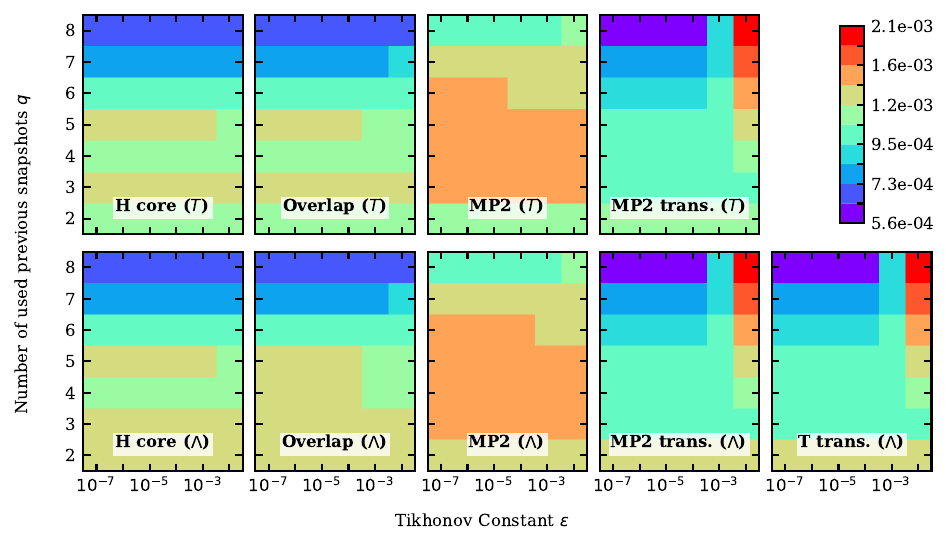}
    \caption{The GMRAE for the CC-amplitudes (T) and $\Lambda$-amplitudes $(\Lambda)$ of \textbf{toluene} for various descriptors, numbers of previous snapshots $q$, and Tikhonov constants $\varepsilon$.}
    \label{fig:heatmaps}
\end{figure*}

\begin{figure*}[t!]
    \centering
    \includegraphics[width=\textwidth]{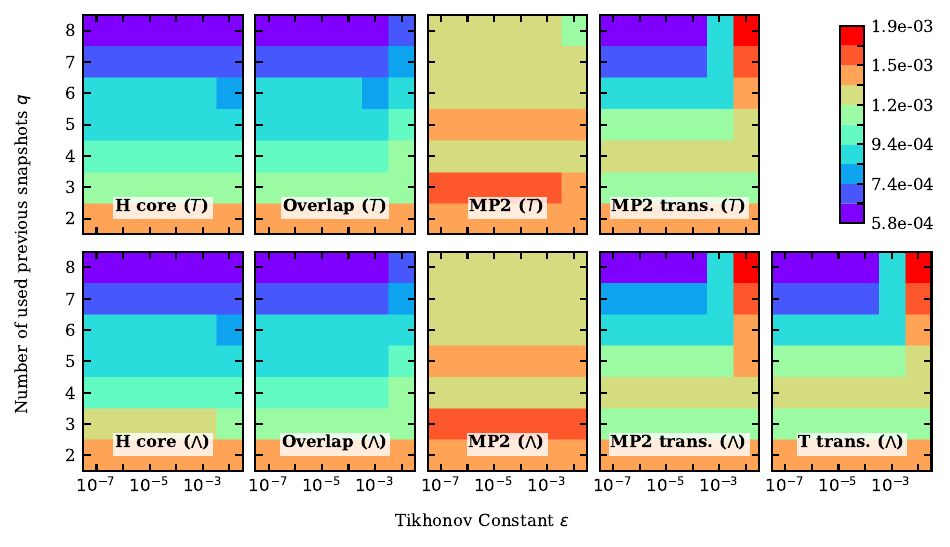}
    \caption{The GMRAE for the CC-amplitudes (T) and $\Lambda$-amplitudes $(\Lambda)$ of \textbf{uracil} for various descriptors, numbers of previous snapshots $q$, and Tikhonov constants $\varepsilon$.}
    \label{fig:heatmaps}
\end{figure*}

\begin{figure*}[t!]
    \centering
    \includegraphics[width=\textwidth]{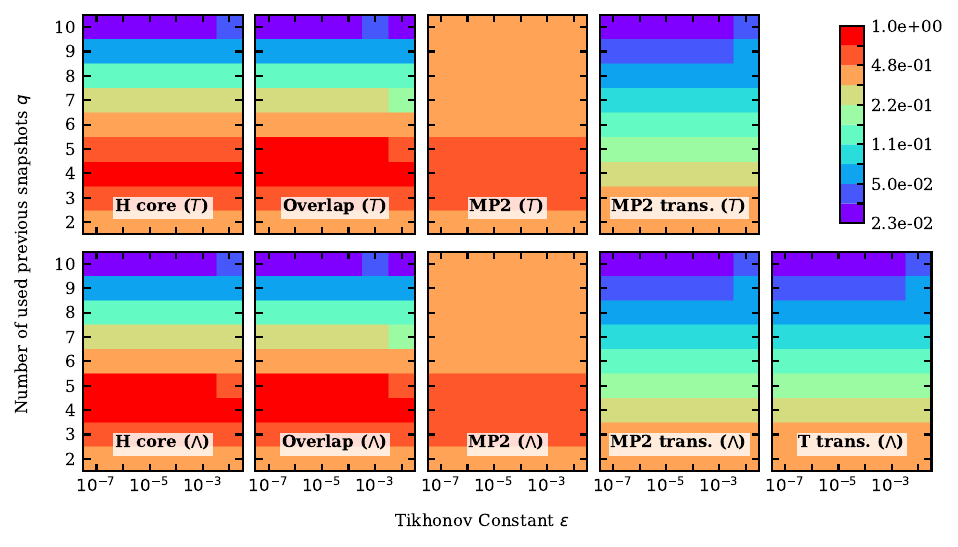}
    \caption{The GMRAE for the CC-amplitudes (T) and $\Lambda$-amplitudes $(\Lambda)$ of \textbf{isobutane} for various descriptors, numbers of previous snapshots $q$, and Tikhonov constants $\varepsilon$.}
    \label{fig:heatmaps}
\end{figure*}

\begin{table*}[t!]
  \centering
  \caption{Deviation of the extrapolated geometry optimisations from the
    reference trajectories. The RMSD is reported for the final geometry and as the
    arithmetic mean over all common optimisation steps; likewise for the
    energy error $\Delta E = E - E^{\mathrm{ref}}$, whose step average is
    taken over $\lvert \Delta E \rvert$.}
  \label{tab:geom-energy-errors}
  \begin{tabular}{l S[table-format=1.3e2] S[table-format=1.3e2]
                    S[table-format=+1.3e2] S[table-format=1.3e2]}
    \toprule
    & \multicolumn{2}{c}{$\mathrm{RMSD}$ / \AA}
    & \multicolumn{2}{c}{$\Delta E$ / kcal\,mol$^{-1}$} \\
    \cmidrule(lr){2-3} \cmidrule(lr){4-5}
    Molecule
    & {final} & {average}
    & {final} & {average} \\
    \midrule
    2-acetyl-1-pyrroline & 1.277e-08 & 8.661e-09 & {0} & {0} \\
    acetone & 1.309e-08 & 8.743e-09 & {0} & {0} \\
    alanine & 2.675e-06 & 1.049e-06 & +6.275e-06 & 2.789e-06 \\
    aspartate & 1.426e-07 & 1.020e-07 & {0} & 1.846e-07 \\
    benzaldehyde & 1.860e-07 & 8.359e-08 & {0} & 2.510e-06 \\
    $\beta$-lactone & 3.883e-08 & 2.662e-08 & 6.275e-06 & 2.092e-06 \\
    cysteine & 4.195e-07 & 1.796e-07 & {0} & 7.470e-08 \\
    cytosine & 7.110e-08 & 2.079e-08 & 6.275e-06 & 4.827e-07 \\
    diacetyl & 1.498e-07 & 7.552e-08 & 6.275e-06 & 3.138e-06 \\
    diallyl~disulfide & 3.883e-07 & 2.422e-07 & 6.275e-06 & 3.922e-07 \\
    ethanol & 1.762e-07 & 4.812e-08 & 1.255e-05 & 9.413e-06 \\
    ethyl~acetate & 4.941e-07 & 2.847e-07 & {0} & {0} \\
    glycerol & 2.128e-06 & 4.557e-07 & {0} & 4.365e-06 \\
    glycine & 8.532e-06 & 9.200e-07 & {0} & 2.789e-06 \\
    histidine & 1.093e-06 & 6.669e-07 & {0} & {0} \\
    isobutane & 2.753e-06 & 3.449e-07 & +1.255e-05 & 4.564e-06 \\
    isoleucine & 4.974e-07 & 1.705e-07 & {0} & {0} \\
    isopropanol & 2.072e-05 & 6.474e-05 & 6.275e-06 & 2.991e-04 \\
    leucine & 1.845e-07 & 1.168e-07 & {0} & 8.964e-07 \\
    menthol & 4.905e-08 & 2.610e-08 & {0} & {0} \\
    methionine & 1.096e-07 & 8.103e-08 & {0} & 4.482e-07 \\
    myrcene & 3.187e-08 & 5.158e-09 & {0} & 1.107e-06 \\
    oxirane & 4.195e-08 & 3.147e-08 & {0} & {0} \\
    pentadiene & 4.302e-08 & 2.696e-08 & 6.275e-06 & 3.586e-06 \\
    proline & 1.391e-06 & 4.579e-07 & {0} & 6.655e-07 \\
    pyrrole & 2.210e-08 & 2.449e-08 & {0} & {0} \\
    serine & 2.932e-06 & 1.634e-06 & +6.275e-06 & 1.345e-06 \\
    sotolon & 1.698e-08 & 8.246e-09 & {0} & 3.922e-07 \\
    toluene & 8.443e-08 & 4.975e-08 & {0} & 2.092e-06 \\
    trimethylamine & 8.367e-08 & 8.129e-08 & +6.275e-06 & 2.282e-06 \\
    uracil & 1.716e-07 & 1.141e-07 & {0} & 6.972e-07 \\
    valine & 2.542e-07 & 1.197e-07 & {0} & 5.491e-07 \\
    \midrule
    mean\textsuperscript{a} & 1.435e-06 & 2.256e-06 & 2.549e-6 & 1.081e-05  \\
    \bottomrule
  \end{tabular}
 
  \vspace{0.5em}
  \footnotesize
  \raggedright
  \textsuperscript{a}\,Arithmetic mean of the absolute values over all
  32 molecules.\\
  Entries of $0$ indicate agreement to the full precision of the printed
  energies ($10^{-8}\,E_{\mathrm{h}}$), i.e.\ $\lvert \Delta E \rvert <
  6.3 \times 10^{-6}$\,kcal\,mol$^{-1}$.
\end{table*}